\documentclass[twocolumn,times]{aastex631}
\usepackage{amssymb}
\received{\today}
\revised{}
\accepted{}
\submitjournal{ApJ}

\usepackage{color,comment}
\usepackage[normalem]{ulem}
\usepackage{newtxtext,newtxmath}
\usepackage{amsmath} 
\usepackage{ctable}

\newcommand{\qpah}[1]{$q_{\rm PAH}$ }
\newcommand{\lpah}[1]{$L_{\rm PAH}$ }
\newcommand{\lfir}[1]{$L_{\rm FIR}$ }

\shorttitle{Stellar Masses at High Redshift}
\shortauthors{Narayanan et al.}

\begin{document}

\title[]{Outshining by Recent Star Formation Prevents the Accurate Measurement of High-$z$ Galaxy Stellar Masses}

\correspondingauthor{Desika Narayanan}
\email{desika.narayanan@ufl.edu}

\author[0000-0002-7064-4309]{Desika Narayanan}
\affil{Department of Astronomy, University of Florida, 211 Bryant Space Sciences Center, Gainesville, FL 32611 USA}
\affiliation{Cosmic Dawn Center (DAWN), Niels Bohr Institute, University of Copenhagen, Jagtvej 128, K{\o}benhavn N, DK-2200, Denmark}
\author[0000-0003-4422-8595]{Sidney Lower}
\affil{Department of Astronomy, University of Florida, 211 Bryant Space Sciences Center, Gainesville, FL 32611 USA}
\author[0000-0002-5653-0786]{Paul Torrey}
\affil{Department of Astronomy, University of Florida, 211 Bryant Space Sciences Center, Gainesville, FL 32611 USA}

\author[0000-0003-2680-005X]{Gabriel Brammer}
\affiliation{Cosmic Dawn Center (DAWN), Niels Bohr Institute, University of Copenhagen, Jagtvej 128, K{\o}benhavn N, DK-2200, Denmark}
\author[0000-0002-2113-4863]{Weiguang Cui}
\affil{Departamento de Física Teórica, M-8, Universidad Autónoma de Madrid, Cantoblanco 28049, Madrid, Spain}
\affil{Centro de Investigación Avanzada en Física Fundamental (CIAFF), Universidad Aut\'{o}noma de Madrid, Cantoblanco, 28049 Madrid, Spain}
\affil{Institute for Astronomy, University of Edinburgh, Royal Observatory, Edinburgh EH9 3HJ, United Kingdom}

\author[0000-0003-2842-9434]{Romeel Dav\'e}
\affil{Institute for Astronomy, University of Edinburgh, Royal Observatory, Edinburgh EH9 3HJ, UK 7Department of Physics and Astronomy}
\affil{University of the Western Cape, Bellville, Cape Town 7535, South Africa}
\author[0000-0001-9298-3523]{Kartheik G. Iyer}
\altaffiliation{Hubble Fellow}
\affiliation{Columbia Astrophysics Laboratory, Columbia University, 550 West 120th Street, New York, NY 10027, USA}
\author[0000-0001-8015-2298]{Qi Li}
\affil{Max Planck Institute for Astrophysics, Garching bei Munchen, Germany}
\author[0000-0001-7964-5933]{Christopher C. Lovell}
\affil{Institute of Cosmology and Gravitation, University of Portsmouth, Burnaby Road, Portsmouth, PO1 3FX, UK}
\affil{Astronomy Centre, University of Sussex, Falmer, Brighton BN1 9QH, UK}
\author[0000-0002-3790-720X]{Laura V. Sales}
\affiliation{Department of Physics and Astronomy, University of California, Riverside, CA, 92521, USA}
\author[0000-0001-6106-5172]{Daniel P. Stark}
\affil{Steward Observatory, University of Arizona, 933 N Cherry Ave, Tucson, AZ 85721, USA}
\author[0000-0003-3816-7028]{Federico Marinacci}
\affiliation{Department of Physics \& Astronomy ``Augusto Righi", University of Bologna, via Gobetti 93/2, 40129 Bologna, Italy}

\author[0000-0001-8593-7692]{Mark Vogelsberger}
\affil{Department of Physics, Kavli Institute for Astrophysics and Space Research, Massachusetts Institute of Technology, Cambridge, MA 021\
39, USA}





\begin{abstract}
We demonstrate that the inference of galaxy stellar
masses via spectral energy distribution (SED) fitting techniques for
galaxies formed in the first billion years after the Big Bang carries
fundamental uncertainties owing to the loss of star formation history (SFH) information from the very first episodes of star formation in the integrated spectra of galaxies.   While this early star formation can contribute substantially to the total stellar mass of high-redshift systems, ongoing star formation at the time of detection outshines the residual light from earlier bursts, hampering the determination of accurate stellar masses.
As a
result, order of magnitude uncertainties in stellar masses can be
expected.  We demonstrate this potential problem via direct numerical simulation of
galaxy formation in a cosmological context.  In detail, we carry out two cosmological simulations
with significantly different stellar feedback models which span a
significant range in star formation history burstiness.  We compute
the mock SEDs for these model galaxies at $z=7$ via $3$D dust radiative
transfer calculations, and then backwards fit these SEDs with {\sc prospector}
SED fitting software.  The uncertainties in derived stellar masses that we find for $z>7$ galaxies motivate the development of new techniques and/or star formation history priors to model early Universe star formation.
  
\end{abstract}




\section{Introduction}
\label{section:introduction}
The most common method for determining the stellar mass of a galaxy is
through ultraviolet-near infrared spectral energy distribution (SED)
modeling.  This technique, first developed by \citet{tinsley68a},
\citet{spinrad71a}, and \citet{faber72a}, models the
expected emission from stellar populations as they evolve over an
assumed star formation history (SFH), with the emission reddened by a
wavelength-dependent dust attenuation curve \citep[see][for
  reviews]{walcher11a,conroy12a,salim20a,pacifici23a}.  This powerful technique is
foundational for our current observational understanding of the cosmic
evolution of galaxy star formation rates and stellar masses
\citep[e.g.][]{madau14a}, and indeed a diverse range of methodologies
for SED fitting have been explored in recent years
\citep{brammer08a,kriek09a,dacunha08a,chevallard16a,iyer17a,carnall18a,johnson21a}.

The assumed form for the model star formation history in SED fitting software is an essential
element in deriving galaxy stellar masses.  Traditional functional
forms for the SFH include constant, exponential declining, burst models, and combinations of these amongst others \citep{conroy13a}.
These parameterized forms for SFH (hereafter, ``parametric'' SFHs)
have parameters describing the (for example) normalization,
$e$-folding time and amplitude of bursts that are varied until a match
is found between the synthetic SED produced by the SPS model and the
observed data.  When a solution is found, the model SFH is then used
to infer the stellar mass of the observed galaxy (typically assuming a fixed metallicity).  Of course, the
assumed form of this SFH can severely impact the modeled stellar mass
\citep{michalowski12a,simha14a,acquaviva15a,iyer17a,carnall18a,carnall19a,dudzeviciute20a}.

More recently, a number of codes have explored the impact of more
flexible so-called ``non-parametric'' forms for the model SFH
\citep[e.g.][]{heavens00a,tojeiro07a,iyer17a,johnson21a}.  Non-parametric SFH models do not
have an explicit functional form as the parametric models, but instead
can vary the amplitude of the SFH over a number of redshift or time
bins in the modeled history of the galaxy.  \citet{iyer18a} validated
the usage of non-parametric SFHs constructed via Gaussian processes by ground-truthing these methods against the the Santa Cruz Semi-Analytic Model (SAM)
and the {\sc mufasa} cosmological simulation \citep{dave16a,dave17a}.
Similarly, \citet{lower20a} demonstrated the efficacy of non
parametric SFH techniques by ground-truthing modeled mock SEDs from
galaxy simulations against their true stellar masses.
\citet{lower20a} found that while traditional parametric forms for the
SFH in SED fitting software had uncertainties at the level $\sim 0.4$ dex, 
non-parametric SFH models reduced these uncertainties to a level of $\sim 0.1$ dex.  \citet{leja19b} found that observed galaxies from
the 3D-HST catalog \citep{brammer12a,skelton14a} are systematically more massive
and older when using non-parametric SFHs as compared to parametric
methods, which bring the observed main sequence in line with
theoretical predictions, potentially alleviating the long-standing
tension between theory and observations in the SFR of galaxies at $z
\approx 2$ at a fixed stellar mass \citep{dave08a,vandokkum08a}.

With the successful launch of the JWST in 2021, observations are
characterizing the physical properties of galaxies at unprecedented
redshifts
\citep[e.g.][]{adams23a,atek23a,castellano22a,donnan23a,finkelstein22a,finkelstein22b,harikane23a,labbe23a,naidu22a,robertson23a}.
Beyond providing constraints on the stellar mass buildup of some of
the earliest galaxies in the Universe, JWST observations of
high-redshift galaxies have been used to demonstrate potential
tensions with the standard $\Lambda$CDM model
\citep[e.g.][]{boylankolchin23a,haslbauer22a,labbe23a,lovell19a,mason23a}.  What
has yet to be explored, however, is the ability of traditional SED
fitting methods to accurately derive stellar masses in these earliest
galaxies, and in particular, the problem of stellar 'outshining'.

The problem of stellar outshining refers to the light from recent
  bursts of star formation overwhelming that of older stellar
  populations, making the inference of their stellar masses difficult.
  The potential impact of outshining has a rich history in the
  observational literature.  \citet{papovich01a} demonstrated an
  uncertainty of $\sim 3-8$ in the derived stellar masses of $z\sim
  2-3$ Lyman Break Galaxies due to uncertainties in the star formation
  history modeling.  This result has been further studied by
  \citet{shapley01a,daddi04a,shapley05a,trager08a,graves10a,sorba18a,tacchella22a,topping22a,whitler23a}
  and \citet{gimenez-arteaga23a}, amongst many other studies. From a
  theoretical standpoint, \citet{maraston10a}, \citet{pforr13a}, and
  \citet{suess22a} constructed mock star formation histories combined
  with stellar population modeling to study the potential impact of
  outshining.

The potential impact of stellar outshining by recent star
  formation in stellar mass inference has yet to be studied in bona
  fide cosmological hydrodynamic galaxy formation simulations. In
this paper, we use numerical simulations of galaxy formation,
combined with post-processed radiative transfer and SED fitting
software, to study the impact of outshining on stellar mass
  estimates in galaxies, with a particular emphasis on high-$z$ galaxy
  detections by JWST.  We demonstrate that standard SED fitting
techniques have a difficult time deriving correct stellar masses in
high-$z$ galaxies owing to outshining by recent bursts of star
formation.  We find that early bursts of star formation can contribute
significantly to the total stellar mass of a galaxy.  By the time the
galaxy is detected at relatively late times ($z \approx 7-10$), the
current ongoing star formation outshines the light from evolved stars,
making the integrated buildup of stellar mass difficult to measure.
We demonstrate this effect using similar techniques to
\citet{lower20a}'s study of the efficacy of non-parametric SFH
modeling: we simulate the formation of galaxies in cosmological
simulations, forward model their mock SEDs using dust radiative
transfer, and then fit these mock SEDs using standard techniques. This
allows us to compare to ground truth from the simulations, and assess
the efficacy of the fitting procedure.

In the remainder of this paper, we expand on these points.  In
\S~\ref{section:methods}, we describe our numerical methods; in
\S~\ref{section:physical_and_luminous_properties}, we describe the
physical and luminous properties of the galaxies that we model; in
\S~\ref{section:sed_fits}, we demonstrate the main issues when SED fitting very young galaxies; in \S~\ref{section:discussion}, we provide discussion, and in \S~\ref{section:summary}, we summarize our main results.

\section{Methods}
\label{section:methods}
\subsection{Summary of Methods}
Our main goal is to create mock SEDs from galaxies formed in a
cosmological simulation, and then fit those SEDs as an observer would
in order to ground-truth SED fitting techniques for $z>7$ galaxies.
To do this, we first simulate the high-redshift evolution of galaxies
by conducting cosmological hydrodynamic galaxy evolution simulations.
We then ``forward model'' the emission from these galaxies by coupling
them with dust radiative transfer in order to generate their mock
SEDs.  With these SEDs in hand, we then ``backward model'' them in
order to derive the inferred galaxy physical properties.  Through this methodology, we determine the
relationship between the inferred stellar masses of our modeled
galaxies from their SEDs, and the true stellar mass.

\subsection{Galaxy Formation Simulations}
\label{section:galaxy_simulations}
We employ two rather different cosmological galaxy formation models in
order to ensure the robustness of our results. The primary
differences in these models as far as this study is concerned are:
\begin{enumerate}
\item The stellar feedback model, which impacts the burstiness of the modeled star formation histories.
  \item The dust model, which impacts the dust content, extinction,
    and attenuation of the emergent light.
\end{enumerate}
In detail, we employ the {\sc simba} cosmological simulation, as
already run by \citet{dave19a}, as well as a newly run simulation
employing the {\sc smuggle} explicit stellar feedback model within the
{\sc arepo} code base.  We briefly describe these models in detail in
turn below.

The {\sc simba} simulation is based on the {\sc gizmo} cosmological
gravity plus hydrodynamic solver \citep{hopkins15a}, and evolves dark
matter and gas elements together, including gravity and pressure
forces.  Gas cools radiatively using the {\sc grackle} library
\citep{smith17a}, including both metal line cooling, as well as
non-equilibrium evolution of primordial elements.  Stars form in
molecular H$_2$ gas following the \citet{krumholz09a} subresolution
prescription for determining the HI and H$_2$ content in a gas
particle, as well as the \citet{kennicutt98a} prescription for the
star formation rate.  Here, a star formation efficiency is manually
set to $\epsilon_* = 0.02$.  The gas itself is artificially
pressurized in order to resolve the Jeans mass as described in
\citet{dave16a}.  The upshot of this is that the star formation
history for a given galaxy tends to occur more smoothly than in models
with an explicit feedback model \citep[and no artificial pressurization of
the ISM;][]{iyer20a}.  Dust is modeled within {\sc simba}
following the algorithms outlined in \citet{li19a}.
Specifically, dust is included as a single sized particle, though can
comprise of multiple species (graphites and silicates).  Dust is
formed in evolved stars \citep{dwek98a}, can grow via metal accretion
\citep{asano13a}, and can be destroyed via thermal
sputtering in hot gas \citep{tsai95a,mckinnon17a,popping17a},
supernova blastwaves, and astration in star-forming
regions.  

In addition to the {\sc simba} simulation, we have run simulation with the {\sc smuggle}
galaxy formation model enabled in the {\sc arepo} code~\citep{springel10a, weinberger20a}.  We refer
the reader to \citet{marinacci19a} for a full description of this
model, and highlight only the key differences from the {\sc simba}
simulation as they pertain to our study.  Star formation occurs only
in gravitationally bound molecular gas \citep{hopkins13e}, and follows a
volumetric \citet{kennicutt98a} law, though with an efficiency
$\epsilon_* = 1$.  This is in contrast to the forced inefficiency of
star formation in {\sc simba} because on long timescales, stellar
feedback in the {\sc smuggle} model self-regulates the star formation
rates to result in the relatively low efficiencies observed in
molecular clouds \citep{marinacci19a}.  Stellar feedback models include supernovae,
radiative feedback, stellar winds, and thermal feedback from HII
regions.  Like {\sc simba}, dust is also included in the {\sc smuggle}
model.  In contrast, however, dust is modeled to include a spectrum of
grain sizes that evolve as the dust grains evolve in the simulation.
Beyond the aforementioned dust processes that are included in {\sc
  simba}, the dust in {\sc smuggle} is allowed to grow in size via
coagulation (sticking together), as well as fragment into smaller
grains via grain-grain shattering collisional processes.  The upshot
from this dust modeling is that the local extinction law is explicitly
computed in a spatially resolved sense in galaxies \citep{li21a}, and
represents therefore a fundamental difference in the forward modeling
of radiative transfer between the {\sc smuggle} and {\sc simba}
models.  We have simulated a $25/h$ Mpc side-length box with periodic
boundary conditions, starting from $z=99$ with initial conditions
generated with {\sc music} \citep{hahn11a}.  We have run the model at
the same mass resolution as the {\sc simba} simulation ($2 \times 512^3$
particles), though the initial conditions are generated with different
random seeds, so the galaxies are not directly mappable from one
simulation to another.  We allow the simulation to evolve to redshift
$z=6$, and restrict our analysis to this redshift range.

  \begin{figure}
  \includegraphics[scale=0.6]{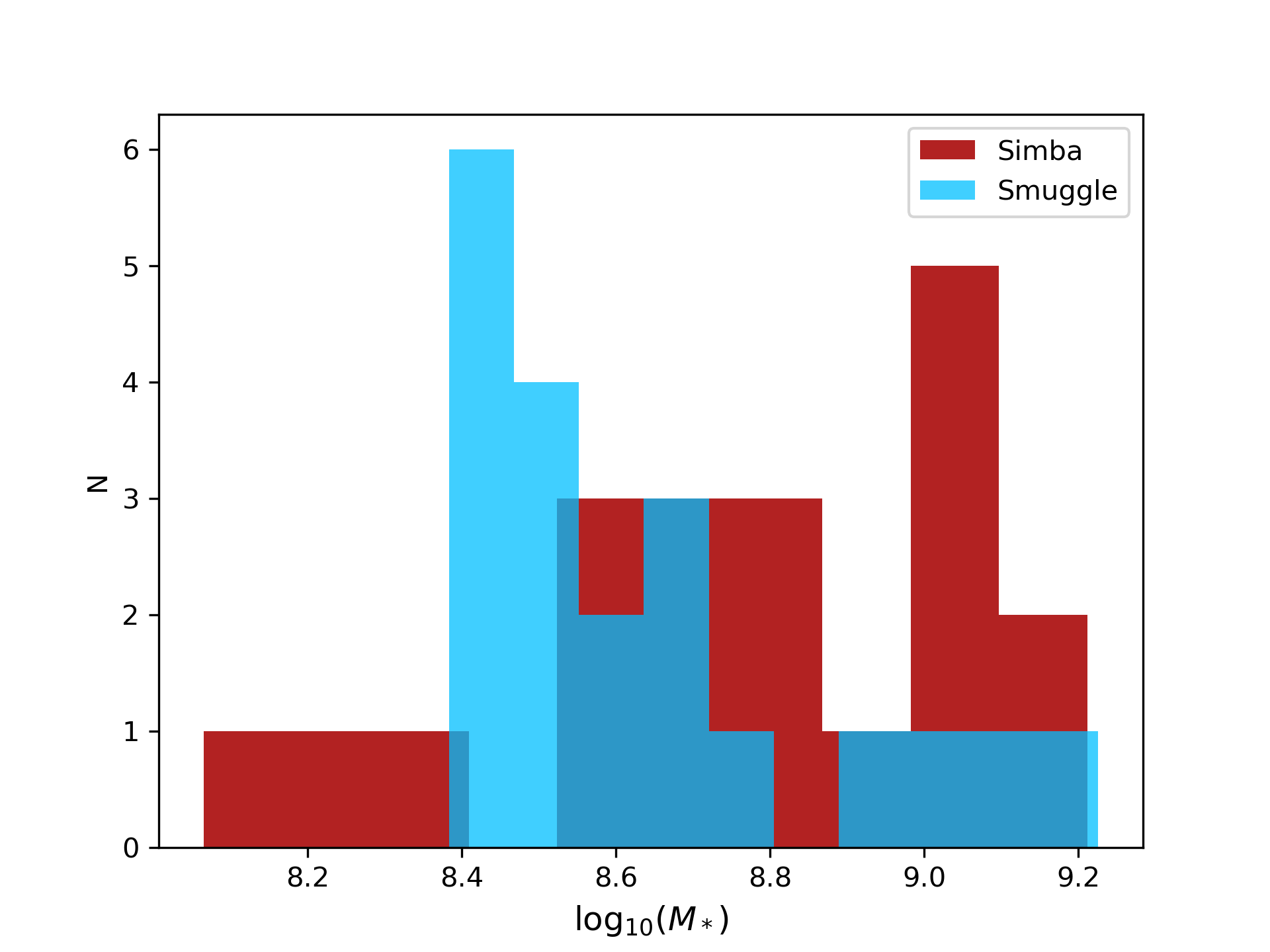}
  \caption{{\bf Histogram of the stellar masses at $z \sim 7.5$ for the
    $20$ most massive galaxies in the {\sc simba} cosmological
    simulation (red) and the {\sc smuggle} cosmological box (blue).}  Both
    boxes are $25/h$ Mpc on a side.\label{figure:mstar_hist}}
\end{figure}

\begin{figure*}
  \includegraphics[scale=0.6]{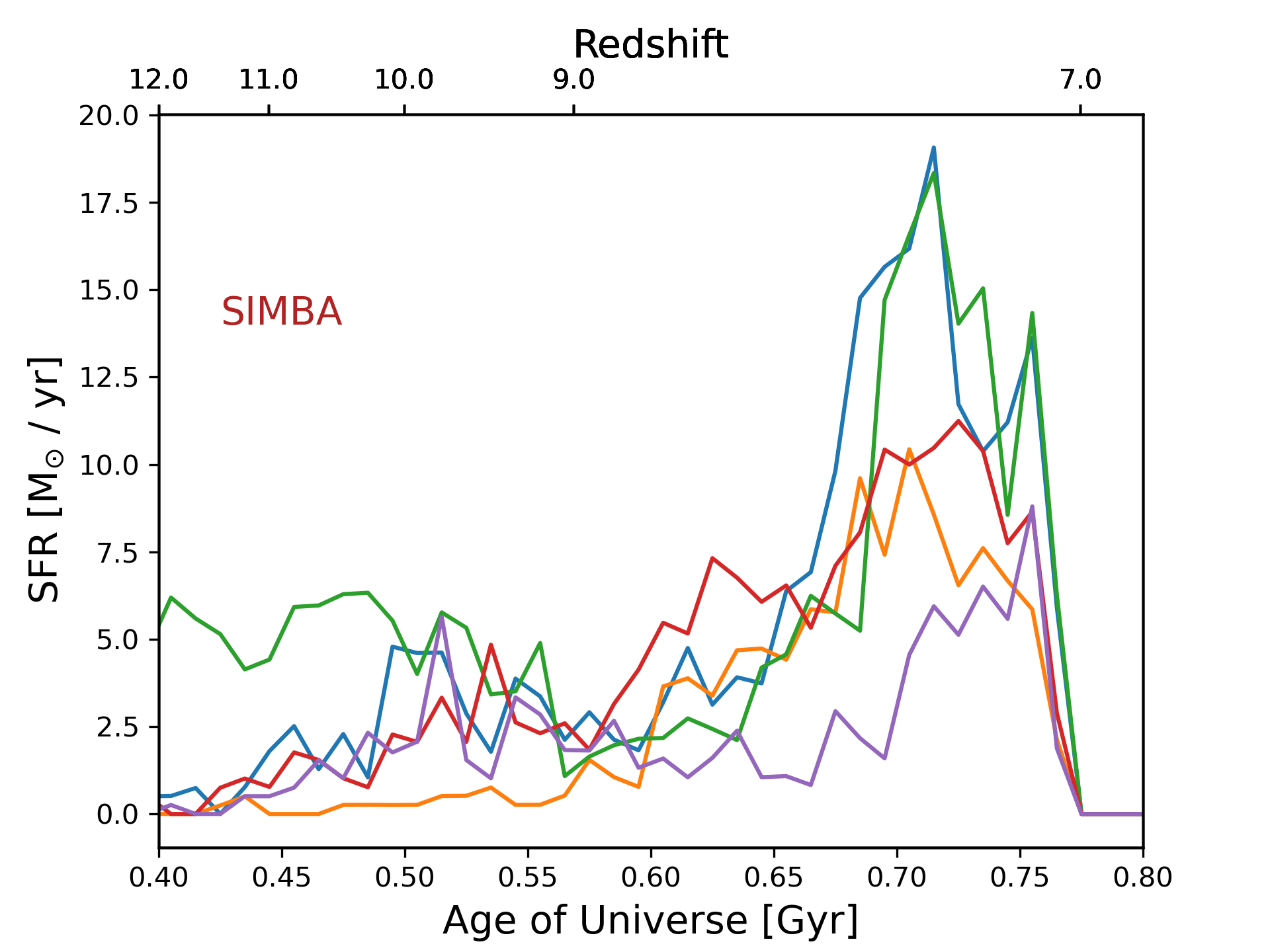}
  \includegraphics[scale=0.6]{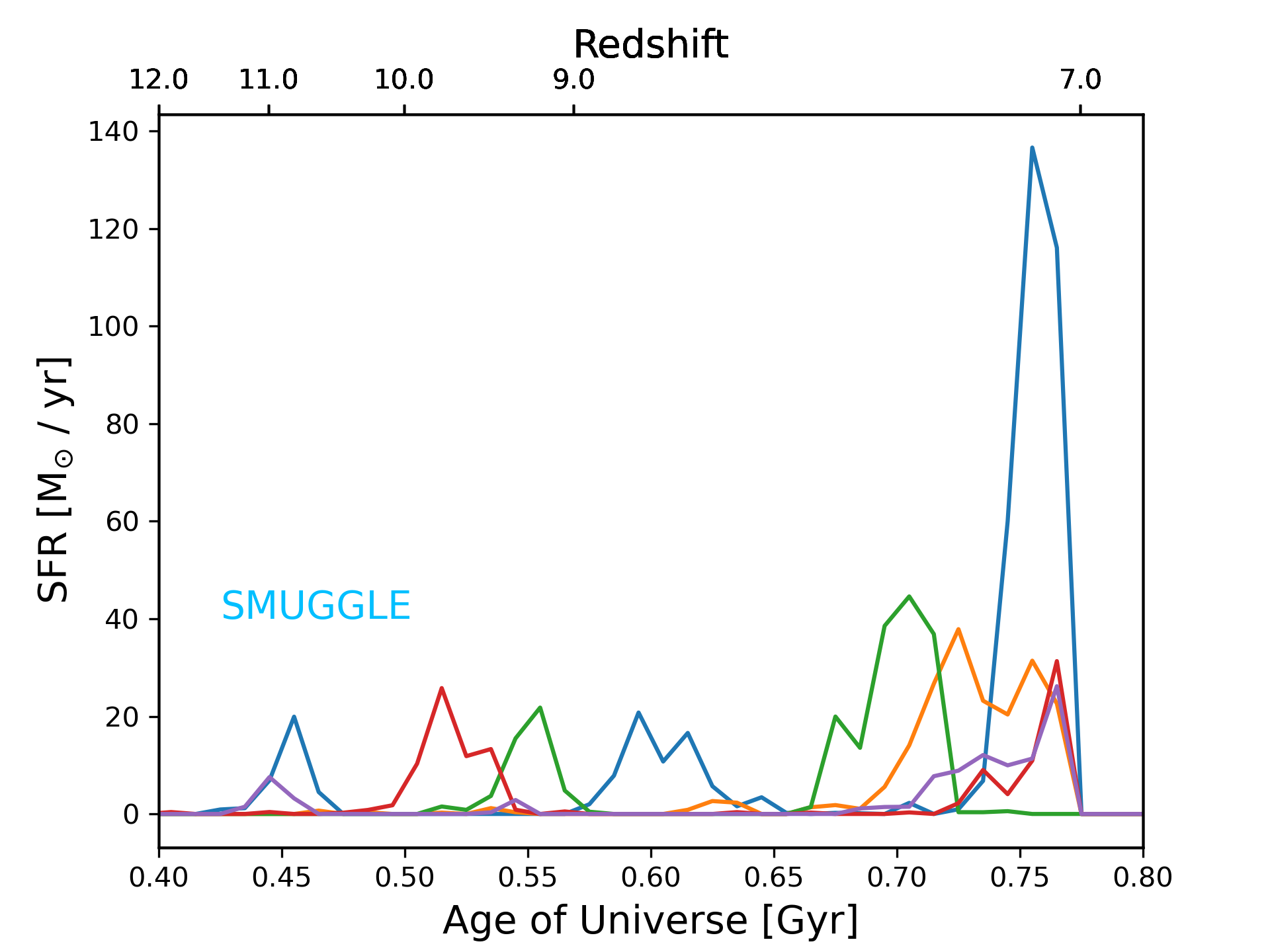}
  \caption{{\bf Example star formation histories of $5$ arbitrarily
      chosen galaxies from the {\sc simba} and {\sc smuggle}
      cosmological boxes.}    The left panel shows {\sc simba}, while the
    right panel shows {\sc smuggle}.  The former has an artificially
    pressurized ISM (as do most cosmological galaxy formation models),
    while the right represents a class of explicit feedback models.  The two models highlight the diversity of
    modeled star formation history burstiness possible in galaxy
    formation models.  We use both of these models in our analysis to
    model a range of possibilities when investigating the efficacy of
    SED fitting in young galaxies.  Note that the sudden drop off in the SFH at $z<7$ is an artifact of our constructing the SFHs at $z=7$, and not physical.} \label{figure:sfh}
\end{figure*}

 \begin{figure*}
    \begin{centering}
  \includegraphics{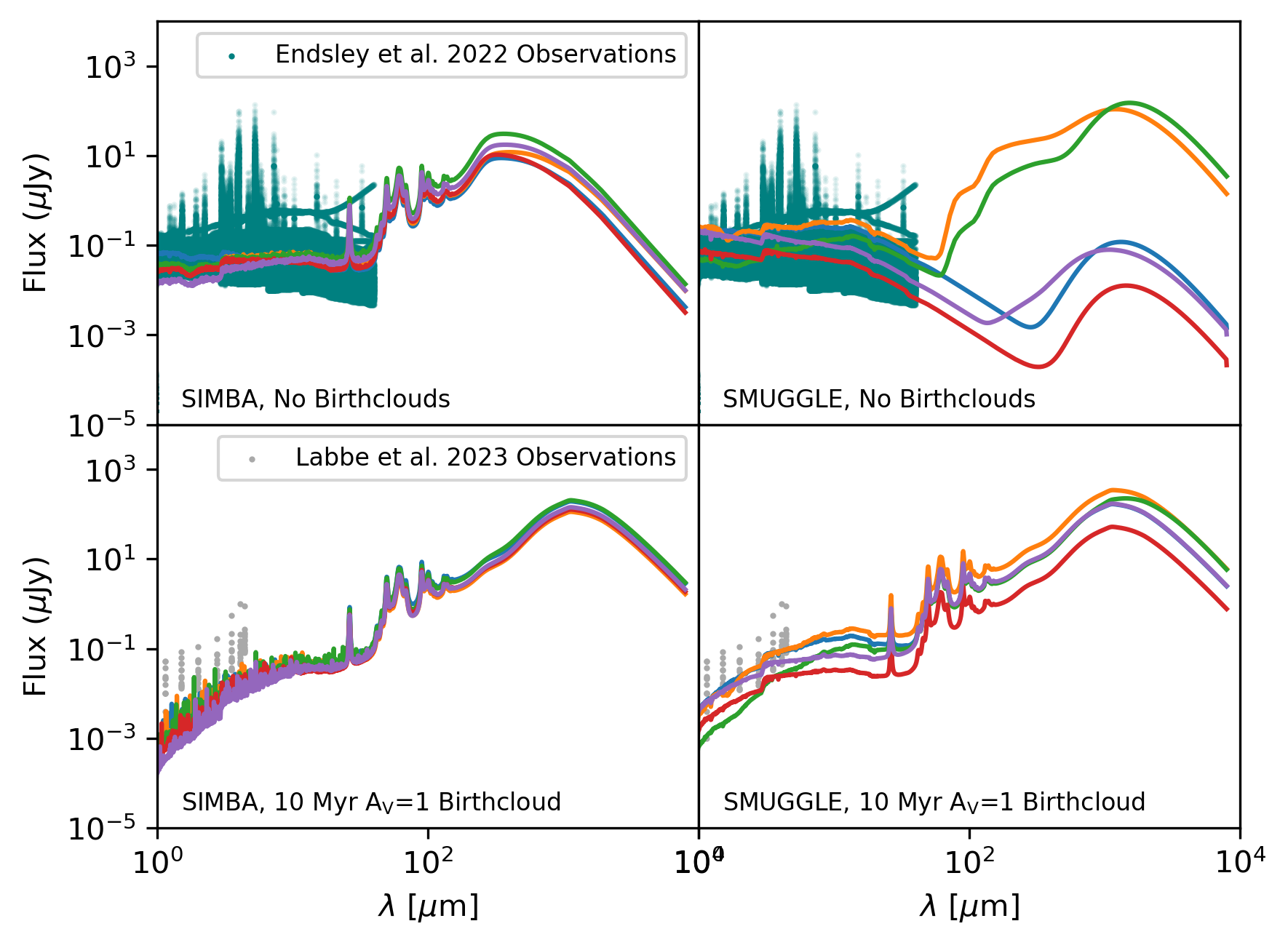}
\caption{{\bf Modeled SEDs for the {\sc simba} (left) and {\sc
      smuggle} (right) models at $z \sim 7.5$.} The top row shows the
  results from {\sc powderday} radiative transfer when only the
  diffuse dust is included, while the bottom row includes
  \citet{charlot00a} birthclouds with an $A_{\rm V} = 1$ screen in
  front of all $t<10$ Myr star particles.  The green points show observations from \citet{endsley22a}, while grey points show observations from \citet{labbe23a}. We find that birthclouds dominate the obscuration in the UV/optical.\label{figure:seds}}
    \end{centering}
    \end{figure*}

\subsection{Dust Radiative Transfer}
In order to generate the mock SEDs (that we will then fit using {\sc
  prospector}), we employ the public {\sc powderday} dust radiative transfer
package \citep{narayanan21a}, which employs {\sc yt, fsps} and {\sc
  hyperion} for grid generation, stellar population synthesis, and
Monte Carlo radiative transfer respectively
\citep{turk11a,conroy09b,robitaille11a}.  Here, stars that form in the
galaxy formation simulation emit a stellar spectrum based on their
ages and metallicities.  This spectrum is computed using {\sc fsps}
\citep{conroy09b,conroy10a,conroy10b}.  We assume the {\sc mist}
stellar isochrones \citep{choi16a}, and a \citet{kroupa02a} stellar
initial mass function (consistent with both sets of hydrodynamic simulations).  
We note that the choices of these parameters can subtly impact our results.  For example, different isochrone models can change the assumed lifetimes of massive stars, which will impact the degree of outshining that we model here.

The light from these stars\footnote{We note that HII regions around massive stars could generate nebular emission lines in the UV/optical, which can possibly impact broadband photometry \citep[e.g.][]{stark13a,gimenez-arteaga23a}.  Indeed, some work has already been done to include nebular lines in {\sc powderday} \citep[e.g.][]{garg22a}, and this emission (once computed) is simply tacked onto the stellar SEDs for the dust radiative transfer phase.  This said, we aim to isolate the uncertainties incurred by the SED fits in our simulations to the star formation histories, and therefore do not include this additional physics in either the forward modeling, or the SED fitting.  Future work will investigate the role of nebular emission in contaminating broad band fluxes in galaxies. } is emitted in an isotropic manner, and can
be absorbed, scattered, and re-emitted by dust in the individual cells
in the galaxy.  For the {\sc simba} simulations, which are particle
based, the dust information is smoothed from the particles onto an
adaptive mesh with an octree memory structure, and the radiative
transfer occurs on this grid.  For {\sc simba} we assume \citet{weingartner01a}
dust extinction laws locally.  For {\sc smuggle}, we
perform the radiative transfer on a Voronoi mesh built around the dust
particles simulated with the active dust model, and compute the extinction laws  explicitly in each cell following \citet{narayanan23a}.
Here, we assume extinction efficiencies from \citet{draine84a} and
\citet{laor93a} for silicates and carbonaceous grains respectively.

Beyond attenuation by the diffuse dust in the galaxy (which is
explicitly modeled in both {\sc simba} and {\sc smuggle}), {\sc
  powderday} includes the possibility of obscuration by subresolution
birth clouds following the \citet{charlot00a} formalism as built into
{\sc fsps}.  Here, the attenuation only occurs for star particles
below a threshold age (we set this to $t=10$ Myr when included), and
has a user-defined normalization to the attenuation curve, that we set
as $A_{\rm V}=1$.  We discuss this in further detail in \S~\ref{section:model_seds}.


\section{The Physical and Luminous Properties of $z>7$ Galaxies}
\label{section:physical_and_luminous_properties}
\subsection{Stellar Masses and Star Formation Histories}
We begin our analysis by describing the physical and luminous
properties of our model galaxies.  
In Figure~\ref{figure:mstar_hist}, we plot a distribution of the
  stellar masses at $z=7$ of the $20$ most massive galaxies in the
  {\sc simba} and {\sc arepo} simulations.  In a ($25/h$)$^3$ volume,
  the most massive galaxies at $z \approx 7.5$ are $M_* \sim 10^8-10^9
  M_\odot$.  These stellar masses are, by and large, built up via a
  series of individual star formation episodes, though the importance
  of these bursts to the total stellar mass buildup (and, in general,
  the shape of the SFH) is dependent on the assumed stellar feedback
  model \citep[e.g.][]{sparre17a,iyer20a}.  In
  Figure~\ref{figure:sfh}, we show the SFHs for $5$ arbitrarily chosen
  galaxies from each simulation.  The SFH is constructed from the $z\approx7$ snapshot, using the $z=7$ stellar ages, correcting for mass loss processes.

  The {\sc simba} model with a manifestly pressurized ISM model
  broadly has a smoother rising star formation history, with
  punctuated bursts superposed.  In comparison, the {\sc smuggle}
  stellar feedback model is dominated by individual bursts at these
  high redshifts, owing to the explicit nature of the feedback model.
  The two models bracket the smoother SFHs typically seen in
  simulations with a pressurized ISM (e.g. {\sc eagle}, {\sc
    illustris}, and {\sc simba}), and the burstier SFHs seen in
  explicit feedback models (e.g. {\sc fire} and {\sc smuggle}).  It is
  the ability to reconstruct these SFHs with reasonably high fidelity
  (or lack thereof) that will prove essential for the SED fitting
  software to accurately derive the stellar masses of galaxies in
  this epoch.

  \subsection{Model SEDs}
\label{section:model_seds}
  In the top row of Figure~\ref{figure:seds}, we show the model SEDs for the same
  arbitrarily chosen galaxies in Figure~\ref{figure:sfh} at $z=7.5$.  As a
  reminder (c.f. \S~\ref{section:galaxy_simulations}), the {\sc simba}
  model actively models the dust content of the galaxies on-the-fly in
  the simulation but does not model the grain size distribution. As a
  result, the interstellar extinction curves are assumed to be those
  of \citet{weingartner01a}.  In contrast, the {\sc smuggle} model
  includes as grain size distribution as well, and therefore the
  radiative transfer used to generate the mock SEDs in
  Figure~\ref{figure:sfh} includes the spatially varying dust
  extinction curves.  The {\sc simba} model uses a Milky Way template for PAH emission (scaled for the local energy deposited), while the {\sc smuggle} model uses the \citet{draine21a} model, based on the local grain populations \citep{narayanan23a}, resulting in fairly different mid-IR emission features.  Therefore, as in modeling the SFHs, the two
  dust models that we include here bracket a reasonable range of modeled
  dust extinction and obscuration.  In the bottom row of Figure~\ref{figure:seds}, we show the same SEDs, but this time include \citet{charlot00a}
  ``birthclouds'' around young stars.  These birthclouds are included in a subresolution fashion such that any star particle with
  age $t_{\rm age} < 10^7$ yr experiences an extra attenuation of
  $A_{\rm V}$=1.    We show observational comparisons from the \citet{endsley22a}\footnote{Note that the data presented here, formally, are the {\sc beagle} SED fits to the data.} and \citet{labbe23a} surveys in the top and bottom, respectively (these are split in the top and bottom for clarity).


Two immediate points are clear from Figure~\ref{figure:seds}.  First, the model SEDs that we present here provide a reasonable match to observations\footnote{Noting, of course, the lack of emission lines in either our forward or backward modeling.  We reiterate that this is intentional so as to isolate the uncertainties modeled here to SFH modeling.}, including SEDs that have relatively blue and relatively red optical spectra.  This allows us to proceed in our analysis with reasonable confidence in our methods.  Second, the observations exhibit a wide range of rest-frame optical colors, with (as a general statement) the \citet{labbe23a} sources being redder than the \citet{endsley22a} galaxies.  When comparing against our models, it is immediately evident that the majority of reddening in $z>7$ galaxies is due to local obscuration at the sites of very young stars; diffuse dust in the ISM is insufficient to provide the required reddening to match the observed NIRCAM photometry for the reddest sources.   This latter point is a net win for galaxy SED fitting: many modern SED fitting codes have the ability to include such clouds in their backwards modeling.  This reduces the
  uncertainties incurred by the diverse shapes of diffuse ISM attenuation curves \citep{narayanan18b,trayford20a,lower22a}.

\section{Recovering the Stellar Masses of redshift $> 7$  Galaxies via SED Fitting}
\label{section:sed_fits}

      \begin{figure}
        \begin{centering}
\includegraphics[scale=0.55]{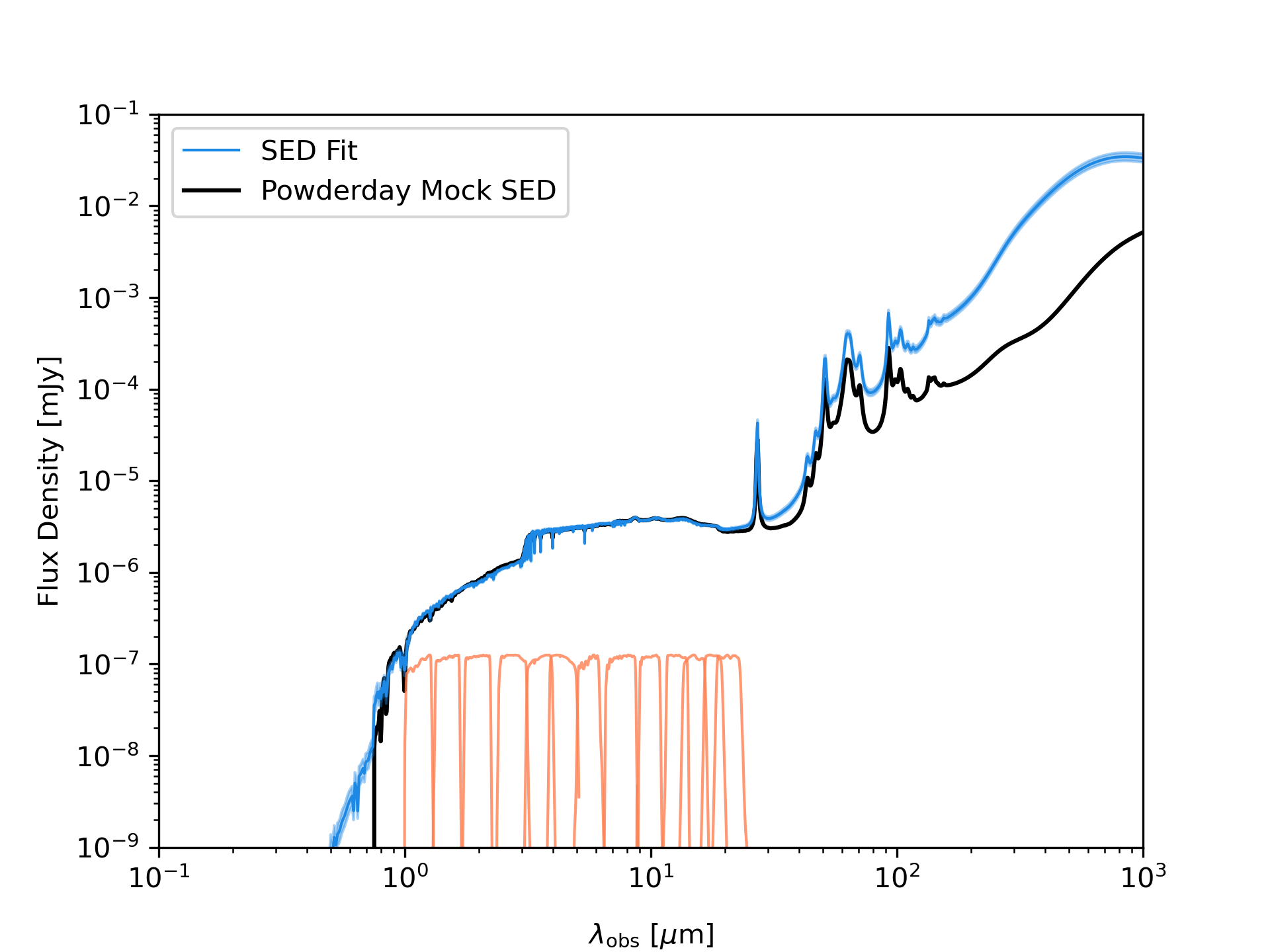}
\caption{{\bf Comparison of SED fit to input SED.}  The input mock galaxy SED generated by {\sc powderday} is in black, while the $16-84^{\rm th}$ percentile confidence intervals in the fit are denoted by the blue shaded region.  The orange lines show the NIRCAM and MIRI filter locations for the fit, and the SEDs are redshifted to $z\approx7$.    \label{figure:pd_prospector}}
        \end{centering}
        \end{figure}
      
      \begin{figure*}
        \begin{centering}
\includegraphics{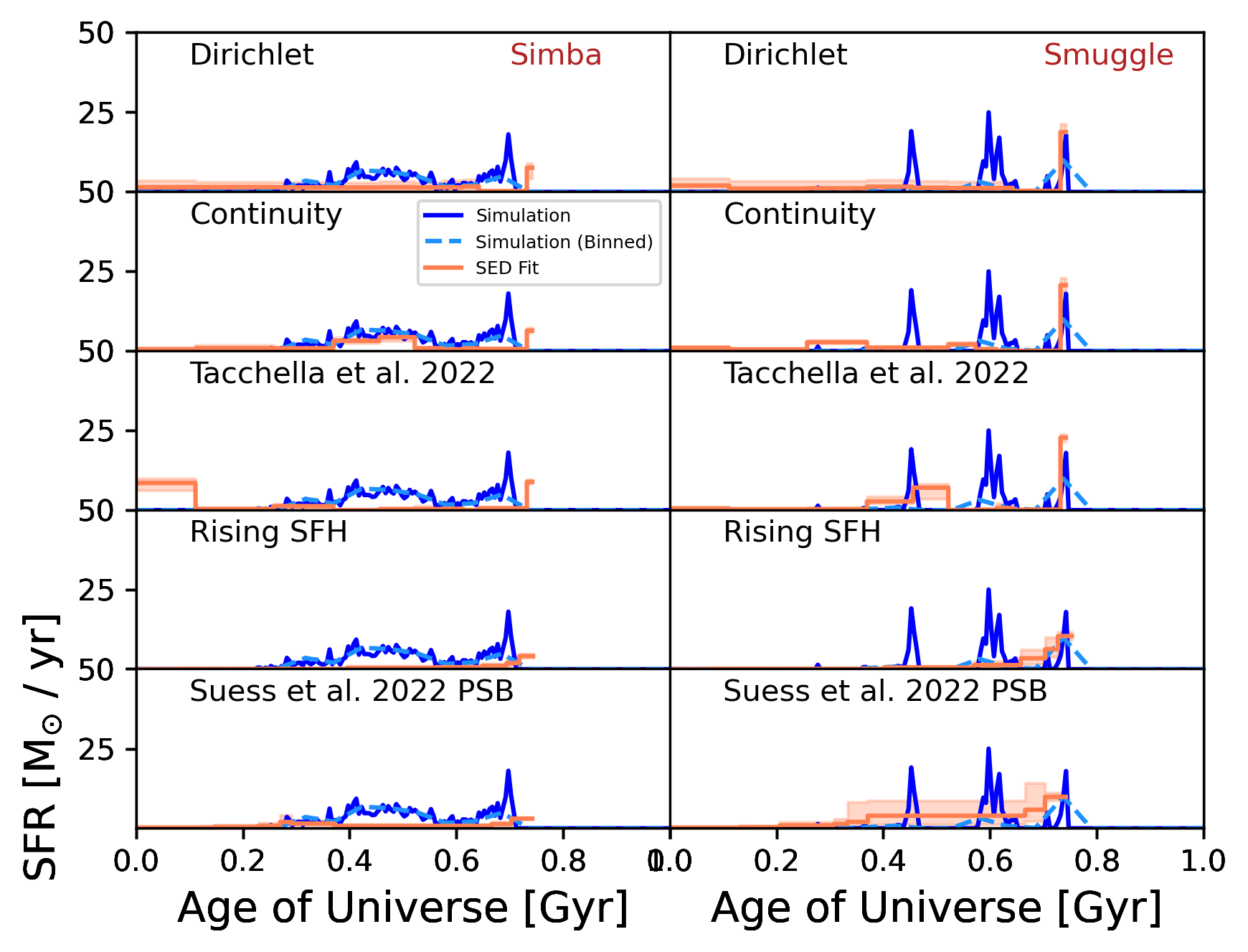}
\caption{{\bf Star formation histories for an arbitrarily chosen galaxies
  from the {\sc simba} simulation (left), and {\sc smuggle} simulation
  (right), with the best fit derived star formation histories from SED
  modeling overlaid (orange).}  The simulated SFH binned at the same time resolution as the SED fits is shown in a lighter, dashed-blue line.   Owing to outshining
  by recently formed stellar populations, early stellar mass buildup can be missed by SED modeling.  This will have a
  significant impact in the derived stellar masses
  (c.f. Figure~\ref{figure:mstar_one2one}). \label{figure:sfh_model}}
        \end{centering}
        \end{figure*}

      \begin{figure*}
        \begin{centering}
\includegraphics{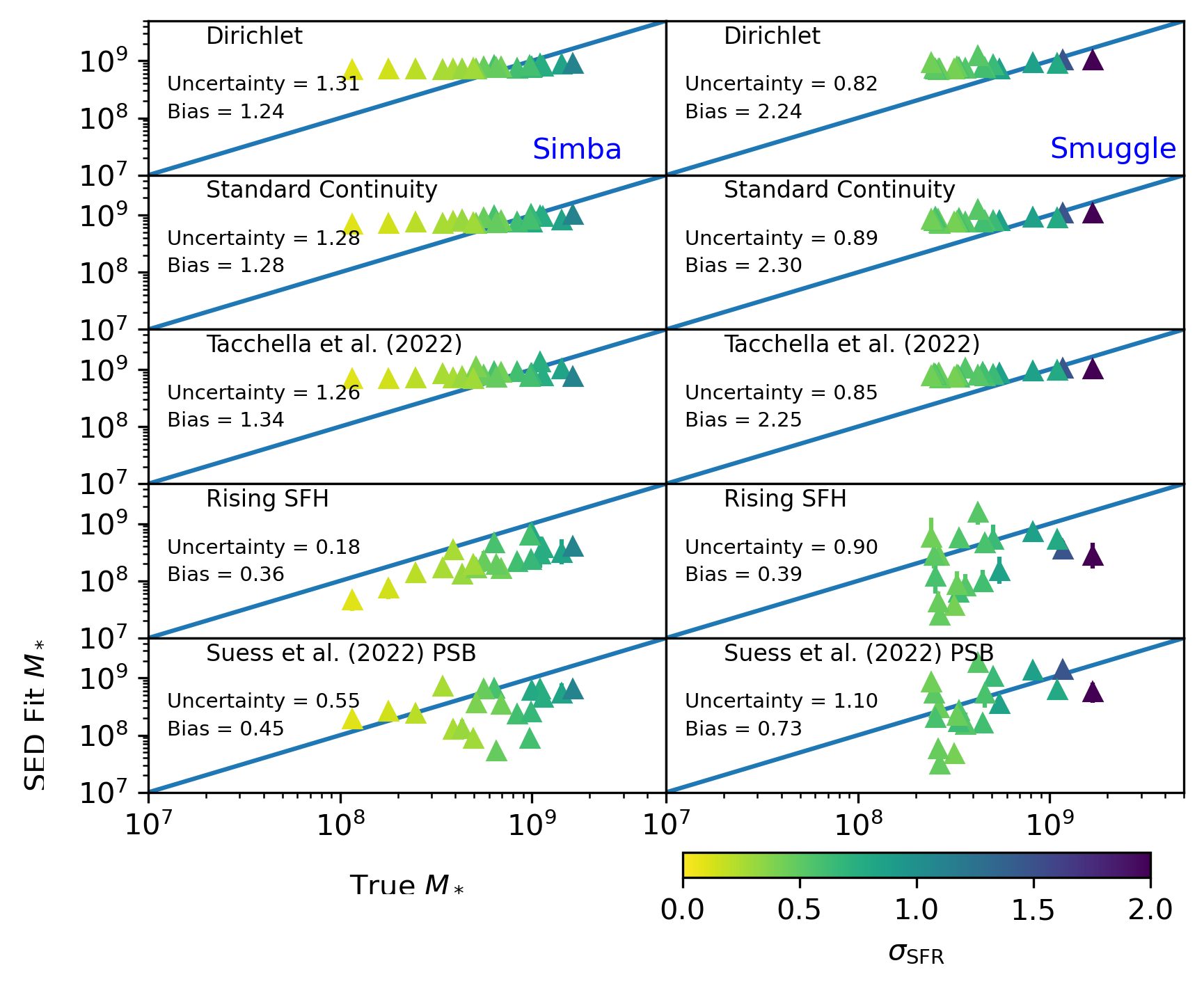}
\caption{{\bf Comparison of the best fit derived $M_*$ from SED
    fitting to the true $M_*$ for the {\sc simba} model (left) and the
    {\sc smuggle model} (right).}  The result of poorly fit star
  formation histories (Figure~\ref{figure:sfh_model}) results in up to
  $\sim$order of magnitude uncertainties in the derived stellar masses
  of high-$z$ galaxies with systematic trends in over(under)
  estimating the $M_*$ depending on the nature of the exact prior
  used.   The color coding shows $\sigma_{\rm SFR}$ which is the standard deviation in the star formation rates over the lifetime of the galaxy, and is a measure of the 'burstiness' of the system. 
 The uncertainty is measured as the standard deviation of $M_*$(fit)/$M_*$(true), while the bias is measured as the median of $M_*$(fit)/$M_*$(true). 
 Generally, the {\sc simba} simulation exhibits less burstiness (c.f. Figure~\ref{figure:sfh}).  The uncertainties and biases vary dramatically, though the priors that favor rising SFHs \citep[e.g.][]{wang23a}, as well as those that allow for more dramatic SFH variations \citep[e.g.][]{suess22a} tend to perform the best.    \label{figure:mstar_one2one}}
        \end{centering}
        \end{figure*}

Having established the nature of early universe star formation histories (at least within the context of two reasonably plausible galaxy formation models), as well as the forward modeled SEDs from these galaxies, we
now ask how accurately we can recover the stellar masses of these
model galaxies via SED fitting the mock SEDs.  We fit these SEDs with
{\sc prospector} \citep{johnson19a}, assuming full JWST NIRCAM and
MIRI coverage of the rest-frame UV-NIR SEDs (noting this is assuming a relatively optimistic wavelength coverage).

By using {\sc prospector} for the backwards modeling, we are able to
minimize the uncertainty incurred by many of the stellar population
parameter choices as fundamentally {\sc prospector} and {\sc
  powderday} both use {\sc fsps} under the hood to model stellar
populations.  We therefore assume the {\it exact same} model IMF,
spectral libraries, and stellar isochrone models in order to obviate
these potential uncertainties in our SED modeling.  We additionally fix the redshift to the true redshift of the galaxy to avoid uncertainties in the redshift fit.  We assume a uniform (and arbitrarily chosen) signal to noise ratio of $10$ across all bands. We allow the birthcloud model in the SED fits to be flexible.

We assume the non-parametric form for star formation history modeling
where the star formation histories are constrained with a set of $3$
model priors:
\begin{enumerate}
  \item The {\bf Dirichlet prior} \citep{betancourt13a} is parameterized by
    concentration index $\alpha$ that sets the preference for all
    stellar mass to be formed in a single bin vs a smoother
    distribution of stellar mass formed over the modeled time period.  We have run tests ranging $\alpha=[0.3,1.0]$ and found minimal impacts on our results.

  \item The {\bf Standard Continuity prior} fits for
    $\Delta$log(SFR) between SFH time bins, thus weighting against
    dramatic variations in the SFH between adjacent time bins.  This
    weighting is parameterized by a Student's t-distribution
    \citep{leja19a}.  This is given by $x={\rm log\left(SFR_n/SFR_{n+1}\right)}$, where:
    \begin{equation}
      {\rm PDF\left(x,\nu,\sigma \right)} = \frac{\Gamma\left(\frac{\nu+1}{2}\right)}{\sqrt{\nu\pi}\Gamma\left(\frac{1}{2}\nu\right)}\left(1+\frac{\left(x/\sigma\right)^2}{\nu}\right)^{-\left(\nu+1\right)/2}
    \end{equation}
Here, $\Gamma$ is the Gamma function, $\sigma$ controls the width
  of the distribution, and $\nu$ controls the tail of the distribution
  function. Following \citet{leja19a}, we set $\nu=2,\sigma=3$.
    \item The {\bf Bursty Continuity prior}, developed by \citet{tacchella22a} adjusts the parameters in the Student's t-distribution to $\sigma=1$ and $\nu=2$, resulting in a burstier SFH.
    
  \item {\bf Rising SFH prior} developed by \citet{wang23a} to
    preferentially favor rising star formation histories.

    \item {\bf PSB prior} developed by \citet{suess22a} for post
      starburst galaxies which have sharp changes in their recent
      SFHs.
\end{enumerate}
These SFH priors are not meant to be comprehensive, but rather to span
a range of reasonable priors commonly used in the literature, and to
demonstrate the potential impact of these priors on the derivation of galaxy stellar masses in the early Universe. 

      In Figure~\ref{figure:pd_prospector}, we show the example fit for one of our model SEDs, with the observational filters overlaid (we zoom in to the NIRCAM/MIRI wavelengths).    The blue shaded region denotes the $16-84\%$ percentile confidence intervals in the posterior, while the black line shows the input {\sc powderday} mock SED.  The fit was performed at $z\approx7$.  The quality of fit presented in Figure~\ref{figure:pd_prospector} is comparable to all of the fits performed for this study. 
      
      In Figure~\ref{figure:sfh_model}, we show the model SFH for an
      arbitrarily chosen galaxy from both the {\sc simba} and {\sc
        smuggle} simulations with the best fit star formation history
      as derived from the {\sc prospector} fits with each of these
      priors imposed (we show this same plot for all modeled
        galaxies in the Appendix).  In the left column we show the
      results for an arbitrarily chosen galaxy from the {\sc simba}
      simulation, while in the right we show the results from the {\sc
        smuggle} model.  The orange lines show the median best fit SFH
      from {\sc prospector}, while the shaded region shows the
      inter-quartile dispersion.  The simulated SFH is shown in solid blue, and in dashed lighter blue, we bin the simulated SFH at the  same time resolution.  As is clear none of the imposed SFH
      priors adequately reproduces the early Universe SFH in either
      galaxy formation model, though the models that allow for the
      most dramatic star SFH variations performs the best.

      In detail, the Dirichlet prior constrains the fraction of
        mass formed in each time bin (in the model SFH) such that the
        fractional sSFR in each time bin follows the Dirichlet
        distribution.  While some variation in the fractional
        contribution to the total stellar mass from a given SFH bin is
        allowed, these do not tend to be dramatic with the Dirichlet prior
        \citep{leja17a}.  As a result, the early SFH for each model is
        relatively low-level, constant SFR, with the dominant changes
        happening in the latest time bins, where the majority of the
        light contributes to the SED.  The Continuity prior is similar
        to the Dirichlet one, though as discussed, favors less dramatic variations in the SFH.
        The default Continuity prior was tuned by \citet{leja19a} to
        the Illustris-TNG star formation histories which are
        relatively smooth (owing to the effective equation of state
        pressurizing the interstellar medium).  As a result, the
        default Continuity prior (as we employ here) will therefore
        demonstrate even less power on short timescales
        \citep{leja19a,lower20a}.  The \citet{tacchella22a} prior
        modifies this Continuity prior to allow for a burstier SFH, though we still find that these
        modifications are insufficient to result in the level of
        burstiness seen in the {\sc simba} and {\sc smuggle}
        simulations.  Because of this, the low level nearly constant
        SFH at early times dominates the stellar mass buildup with the
        Dirichlet, Continuity, and \citet{tacchella22a} modifications
        to the Continuity priors, giving somewhat similar derived
        stellar masses for all model galaxies.

      The rising SFH, developed by \citet{wang23a}, favors rising
        star formation histories.  Generally, the SFHs for both
        the {\sc simba} and {\sc smuggle} models rise with time, with
        the burstiness of the SFH as the main variant between the two.
        As a result, the \citet{wang23a} rising SFH prior performs
        quite well on the {\sc simba} galaxy formation model, which
        have relatively minor bursts, and moderately well on the
        burstier {\sc smuggle} SFH.  Finally, the \citet{suess22a} SFH
        prior is a $2$-component prior designed to fit the SFH of post
        starburst galaxies: in particular, it allows for sharp changes
        in the recent SFH.  As a result, while the \citet{suess22a}
        prior is designed for post starburst galaxies, it performs
        reasonably well for our model high-$z$ galaxies owing to their
        highly time-variable SFHs.

      We show the impact of these SED fits on the derived stellar
      masses in Figure~\ref{figure:mstar_one2one}, where we compare
      the SED fit $M_*$ to the true $M_*$ for each galaxy, with panels
      ordered akin to Figure~\ref{figure:sfh_model}.   We color code the galaxies with a measure of their SFH `burstiness', here parameterized as the standard deviation of the SFR over the history of the Universe.  We additionally provide summary statistics for each model via the uncertainty (quantified as the standard deviation of $M_*$(fit)/$M_*$(true)), and the bias (quantified as the median of the same ratio).  Smaller uncertainties, and biases that are closer to unity demonstrate higher accuracy in the derived stellar masses.
      
      Generally, no
      model performs particularly well, with uncertainties including
      both over-estimates and under-estimates.  The lack of early star
      formation in the SED fits owes to `outshining': the most recent
      burst of star formation dominates the SED, and therefore has a significantly outsized impact on the resulting fit.  The SFH priors that allow for rising SFHs, as well as those that allow for  the most dramatic variations in the SFH quantitatively perform the best (bottom two rows of Figure~\ref{figure:mstar_one2one}), though neither fully capture the very recent star formation variability.

\begin{figure}
        \includegraphics[width=0.48\textwidth]{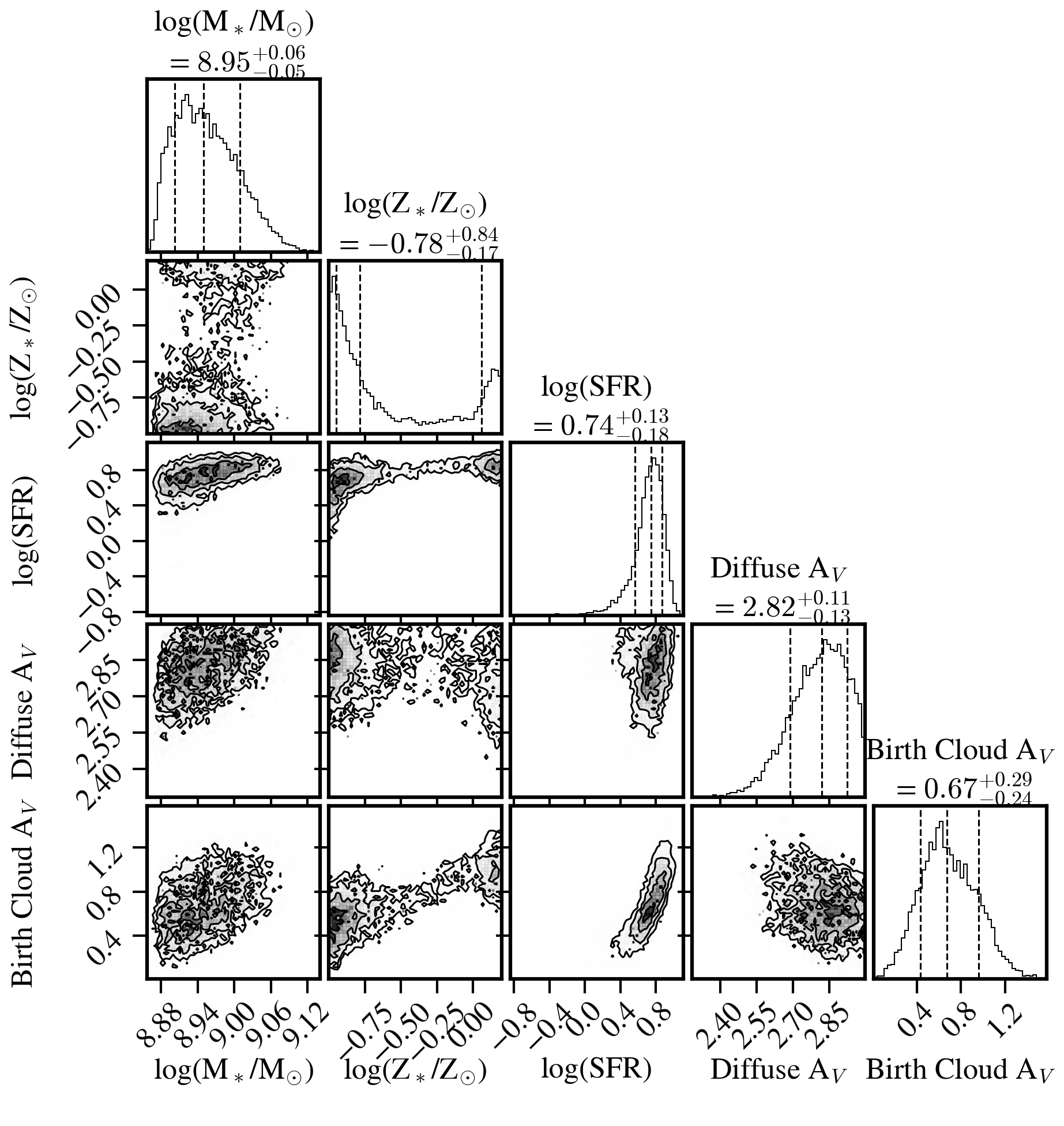}
        \caption{{\bf Example {\sc prospector} model parameter posteriors for an arbitrarily chosen {\sc simba} galaxy fit with the Dirichlet prior SFH model.}  The true physical quantities are: log(M$_*$/M$_\odot$)=9.2; log(SFR/M$_\odot$ yr$^{-1}$)=0.85; log(Z/Z$_\odot$)) = -1.8;  The stellar mass is moderately degenerate with the dust attenuation parameters while the SFR is highly degenerate with the birth cloud dust component. The stellar metallicity posterior is bimodal, so is not well constrained by the data.\label{figure:corner}}
      \end{figure}

       Whether an SED fit over-predicts or under-predicts the
      true stellar mass depends in large part on the amount of early
      star formation that the SED fit predicts.  In
      Figure~\ref{figure:corner}, we show the corner plot for an
      arbitrarily chosen {\sc simba} galaxy using the Dirichlet prior.  There are significant uncertainties in the derived
      physical properties as well as co-variances between them, specifically between stellar mass and SFR and the dust attenuation parameters. These degenerate solutions make inferring the true galaxy properties difficult when the available data does not have enough constraining power.

      We caution that these results are particular to the bands
      employed here (NIRCAM+MIRI), as well as the particulars of the
      SFHs modeled, and are therefore intended to demonstrate the
      range of uncertainty rather than the specific direction of
      uncertainty in $M_*$ estimates at high-$z$.  The take-away from
      this analysis is not that a particular SFH prior tends to under
      predict or over predict stellar masses at high-redshift, but
      rather a demonstration of the relative level of uncertainty in
      fitting the SEDs of galaxies whose stellar masses are dominated
      by early star formation that is being outshined by current star
      formation.

      \section{Discussion}
    \label{section:discussion}

      \subsection{General Discussion}
      We have, thus far, seen that when observing galaxies at high
      redshift (here, we model $z \ga 7$), SED fitting techniques have
      difficulty in correctly inferring the stellar masses of galaxies
      owing to the substantial contribution of stellar mass build up
      by individual star formation episodes at earlier times.  This is true for a range of star formation histories, ranging from the smoother SFHs seen in traditional cosmological simulations to more bursty SFHs in explicit feedback models.  The fundamental issue is that current star formation at the time of detection is able to outshine the prior stellar mass buildup, making it difficult for SED fitting software to infer its presence.  This situation will be most extreme for very bursty systems, though at least within the context of the range of models explored here, appears to be a generic problem.  Whether SFHs at high-$z$ are bursty in the early Universe remains an open question, though recent observations and models appear to suggest at least some level of burstiness \citep[e.g.][]{dressler23a,endsley23a,shen23a}.

      
     As these systems evolve to lower redshift, it is expected
      that traditional SED fitting techniques will perform
      substantially better as the ratio of flux from current star formation to that from the integrated stellar mass buildup  decreases. Indeed, this was demonstrated explicitly by
      \citet{lower20a}, who performed experiments similar to those
      presented in this paper for $z=0$ model galaxies in the {\sc
        simba} simulation, and found that SED fitting with
      non-parametric SFHs models could accurately determine galaxy
      stellar masses. 

      This said, we note some caution when interpreting the results presented here. 
The simulations here encompass two different models for stellar feedback that result in varied SFHs, for galaxies drawn from a relatively small box ($25/h$) Mpc on a side).   While these models span a diverse range of SFHs for similar mass galaxies, they are not necessarily comprehensive.  It is possible that some model forms of star formation history (i.e. where most of the mass is formed at the time of observation) could result in highly accurate derivations of the stellar mass.  Similarly, the modest size of our cosmological boxes excludes the most massive and rare systems.  
These results should not be taken to mean that a particular SFH prior will always overpredict, or always underpredict stellar masses: the exact relationship between the modeled stellar masses of galaxies and true ones depend on a wide range of choices in SED modeling \citep[e.g.][]{conroy13a,lower20a,gilda21a}.  Instead, these results are to simply intended to reflect the uncertainty associated with SED modeling early Universe galaxies.

\subsection{Relationship to other Studies}

      While we have explicitly demonstrated the issues of deriving
      stellar masses in high-$z$ galaxies via direct numerical
      simulation, this issue has been hypothesized in the
      observational literature in a wide range of contexts.  For
      example, \citet{tacchella22a} studied the stellar populations of $11$ galaxies from $9<z<11$, and found that the inferred stellar ages were significantly impacted by the assumed SFH prior, and noted that multiple priors were able to fit the data equivalently.   Similarly, 
      \citet{topping22a} modeled the stellar
      masses of UV selected galaxies at $z \sim 7-8$, and found that the derived
      stellar masses can be uncertain by up to an order of magnitude
      owing to the outshining of older stellar populations by a
      current burst.  \citet{whitler23a} modeled the stellar ages of
      UV-bright $z\sim 7$ galaxies, and found a potential tension
      between the relatively young inferred stellar ages of their
      sample, and the detection rate of higher-redshift sources by
      JWST.  This tension can be alleviated if the redshift $z \sim 7$
      galaxies have older stellar components formed in earlier bursts,
      as in the simulations presented here.  \citet{gimenez-arteaga23a} fit spatially resolved measurements of $5<z<9$ galaxies in the JWST SMACS 0723 field with {\sc bagpipes} \citep{carnall18a}, and found evidence for both a bursty SFH, as well as potential for an older stellar population (outshined by current star formation).  When taking this outshining into account, \citealt{gimenez-arteaga23a} found reduced inferred stellar masses by $0.5-1$ dex.      Finally, \citet{iyer19a}
      note, in their development of the dense-basis methodology for
      non-parametric SFH reconstruction from observed galaxy SEDs,
      that the ability to model older stellar populations is
      prior-dominated rather than likelihood-dominated, and that sharp
      variations in the SFH may be difficult to infer.

      In addition to studying the impact of outshining on stellar
        mass inference of high-$z$ galaxies, we have additionally
        analyzed the impact of star formation history priors on the
        derived masses.  The inference of star formation histories has
        been demonstrated by a number of authors to be strongly
        dependent on the assumed priors for the SFH.  For example,
        \citet{leja19a} have derived stellar masses $0.1-0.3$ dex
        larger using non-parameric star formation histories as opposed
        to traditional parametric models in the $3$D-HST galaxy
        catalog.  Similarly, studying galaxies at $z>7$,
        \citet{tacchella22a} and \citet{whitler23a} found inferred
        stellar ages and masses to be highly sensitive to form of the
        assumed SFH in the SED fitting.  Finally, \citet{suess22a}
        showed that the assumed SFH priors can substantially impact
        the  inferred ages of post starburst galaxies via synthetic stellar population modeling.
      

\subsection{Possible Ways Forward}

The primary outcome of this Letter is to demonstrate the uncertainty of the measurement of high-$z$ stellar masses.    We advocate for investment by the community in the development of methods to reduce the bias and uncertainty in these measurements.

One possibility is through the development of new star formation history priors in SED fitting codes.  Already we have seen in Figure~\ref{figure:mstar_one2one} that priors that allow rapid transitions in star formation history perform reasonably well.  In a similar vein, including information from spectral line features that trace star formation over different time scales may help to quantify the burstiness, and inform the modeling of the SFH \citep[e.g.][]{iyer22a}.

As an alternative to traditional SED fitting techniques, machine learning methods may hold promise.  \citet{gilda21a} demonstrated the efficacy of using mock SEDs from large cosmological simulations as a training set for machine learning-based SED fitting software.  \citet{gilda21a} demonstrated that in some circumstances, these techniques can far outperform traditional SED fitting.

    \section{Summary and Outlook}
      \label{section:summary}
      In this paper, we have employed cosmological galaxy evolution
      simulations in order to investigate the ability for modern SED
      fitting techniques to recover the stellar masses of
      high-redshift galaxies.  Owing to the relatively young ages of
      $z \gtrsim 7$ galaxies, early bursts of star formation
      constitute a significant fraction of the formed mass at the time
      of SED modeling.  As a result, if the stellar light from this early star formation (c.f. Figure~\ref{figure:sfh_model}) is outshined
      by late-time star formation at the time of detection, then the recovered stellar masses
      can be incorrect by nearly an order of magnitude (Figure~\ref{figure:mstar_one2one}).  The impact of
      these uncertainties will decrease at lower redshifts, as the ratio of the light emitted from star formation at the time of detection to the the light from the underlying stellar mass decreases.  As JWST pushes the
      frontier of high-redshift science to increasingly early times,
      we encourage the development of new techniques in order to
      accurately derive the physical properties of these first
      galaxies.

\section*{Acknowledgments}
The authors thank the anonymous referee for a careful, and timely
review of this paper. DN expresses gratitude to Adriano Fontana, Paola
Santini, and the organizors of ``The Growth of Galaxies in the Early
Universe - VIII'', where the idea for this paper was borne out of
discussions at the Bad Moos.  DN additionally thanks the Aspen Center
for Physics which is supported by National Science Foundation grant
PHY-1607611, which is where the original framework for the {\sc
  powderday} code base was developed.  DN thanks Mike Boylan-Kolchin,
Chris Hayward, Chia-Yu Hu, Pavel Kroupa, Justin Spilker, Wren Suess,
and Sandro Tacchella for helpful conversations, as well as Ryan
Endsley for providing data from \citet{endsley22a} for comparison
against our models.  DN and PT were supported by NASA ATP grant
80NSSC22K0716.  WC is supported by the STFC AGP Grant ST/V000594/1 and
the Atracci\'{o}n de Talento Contract no. 2020-T1/TIC-19882 granted by
the Comunidad de Madrid in Spain. He also thanks the Ministerio de
Ciencia e Innovación (Spain) for financial support under Project grant
PID2021-122603NB-C21 and ERC: HORIZON-TMA-MSCA-SE for supporting the
LACEGAL-III project with grant number 101086388. The Cosmic Dawn
Center is funded by the Danish National Research Foundation (DNRF)
under grant \#140



\bibliographystyle{mnras}
\bibliography{./full_refs}


\appendix

Here, we show the model SFH for all of our simulated galaxies, as well
as the {\sc prospector} SED fits (for every modeled SFH prior) for
each galaxy.  The galaxies are ordered by mass (i.e. for an individual
plot, the {\sc simba} galaxy and {\sc smuggle} galaxy each represent
the $N^{\rm th}$ most massive galaxy in the cosmological volume.

\begin{figure*}
  \begin{centering}
    \includegraphics{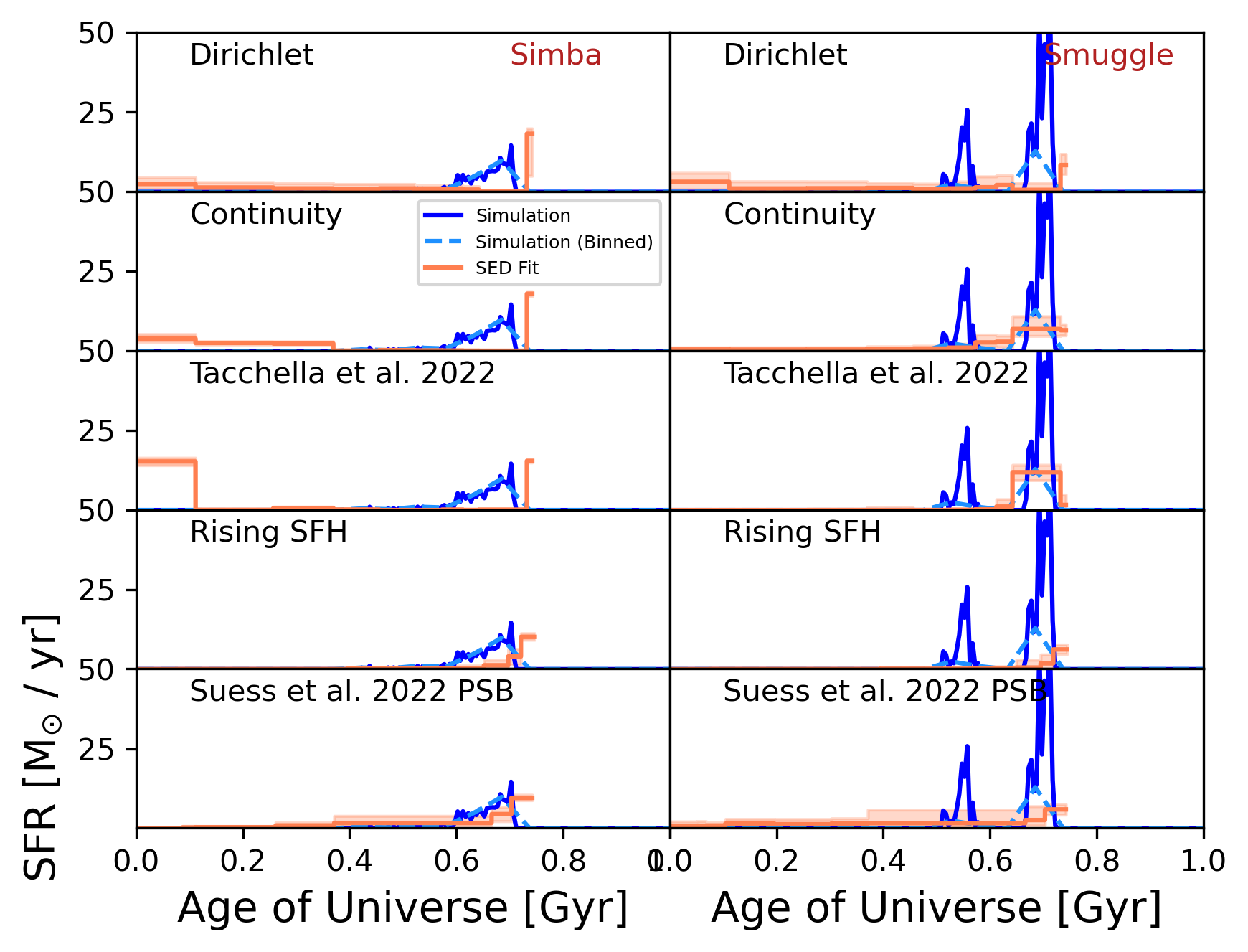}
    \caption{{\bf Star formation histories for the $1^{\rm st}$ most massive galaxy in the {\sc simba} (left) and {\sc smuggle} (right) simulation. }}
  \end{centering}
\end{figure*}

\begin{figure*}
  \begin{centering}
    \includegraphics{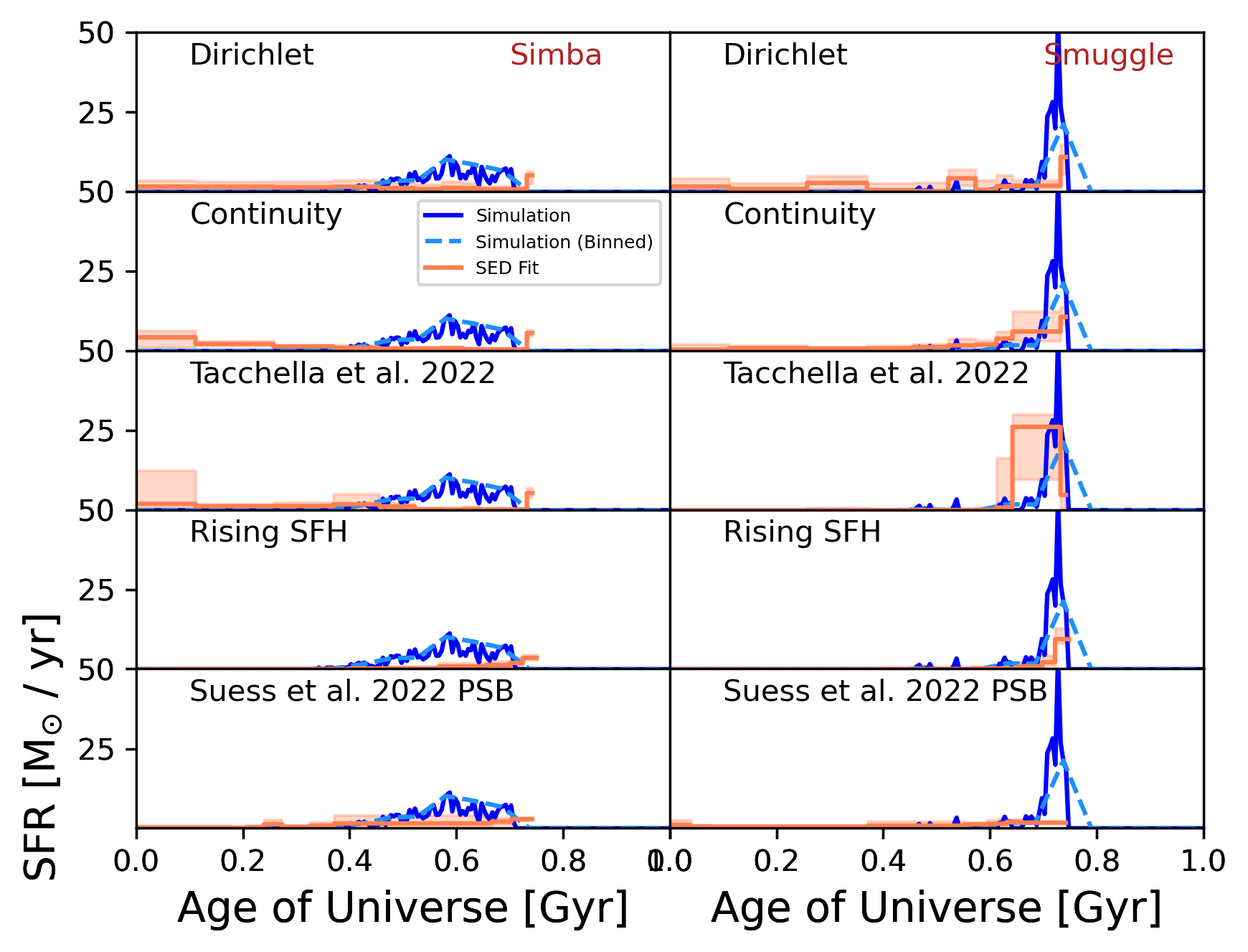}
    \caption{{\bf Star formation histories for the $2^{\rm nd}$ most massive galaxy in the {\sc simba} (left) and {\sc smuggle} (right) simulation. }}
  \end{centering}
\end{figure*}

\begin{figure*}
  \begin{centering}
    \includegraphics{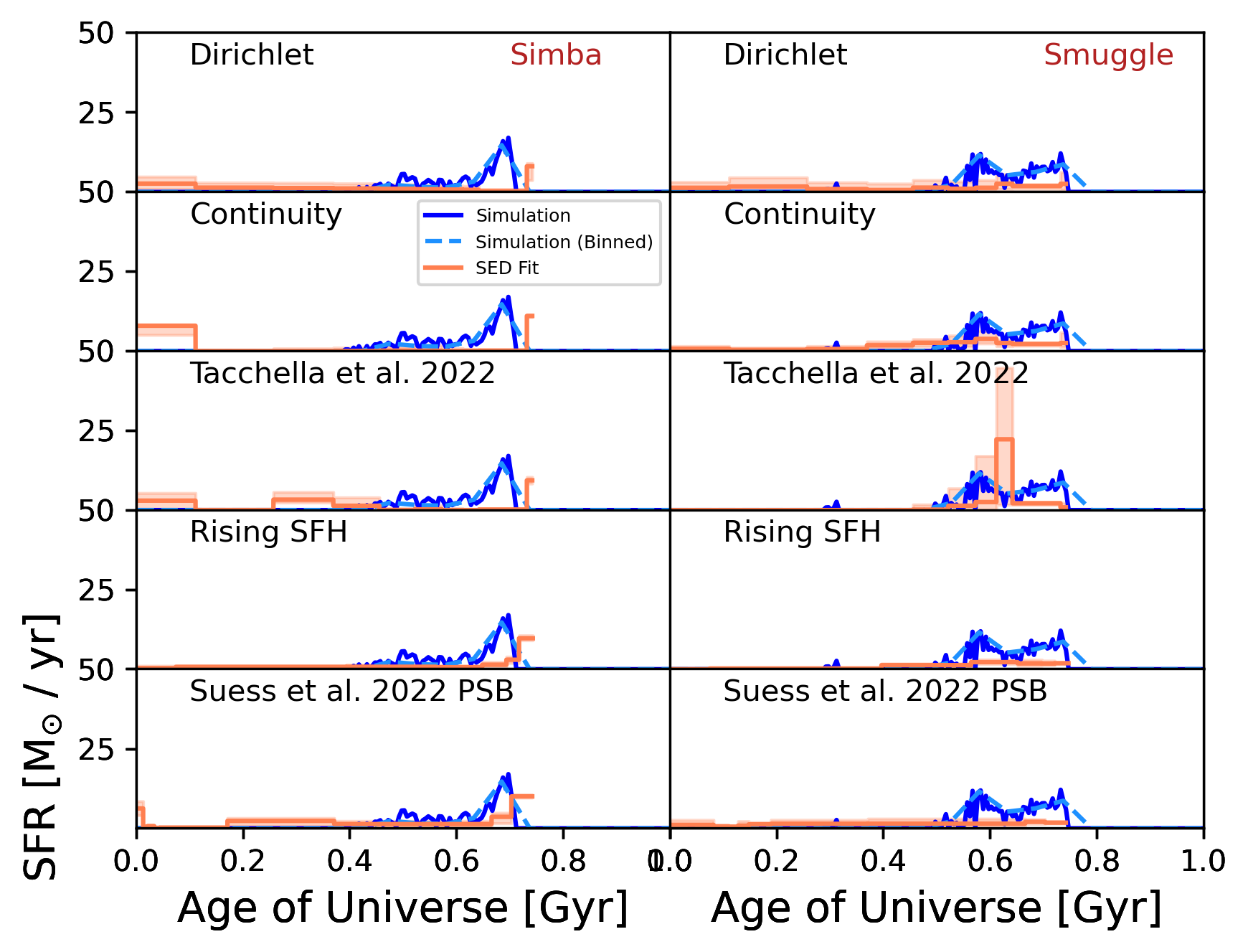}
    \caption{{\bf Star formation histories for the $3^{\rm rd}$ most massive galaxy in the {\sc simba} (left) and {\sc smuggle} (right) simulation. }}
  \end{centering}
\end{figure*}

\begin{figure*}
  \begin{centering}
    \includegraphics{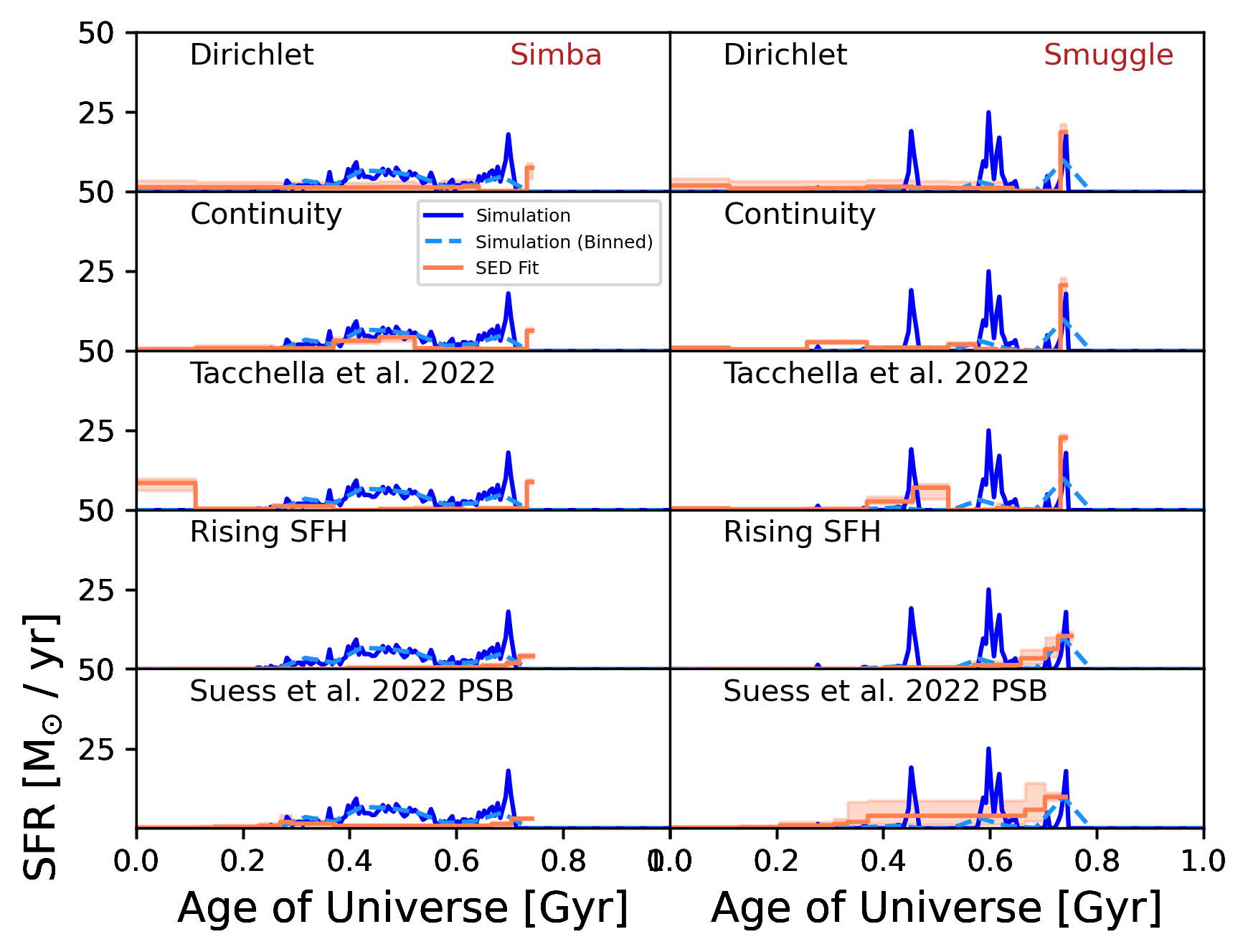}
    \caption{{\bf Star formation histories for the $4^{\rm th}$ most massive galaxy in the {\sc simba} (left) and {\sc smuggle} (right) simulation. }}
  \end{centering}
\end{figure*}

\begin{figure*}
  \begin{centering}
    \includegraphics{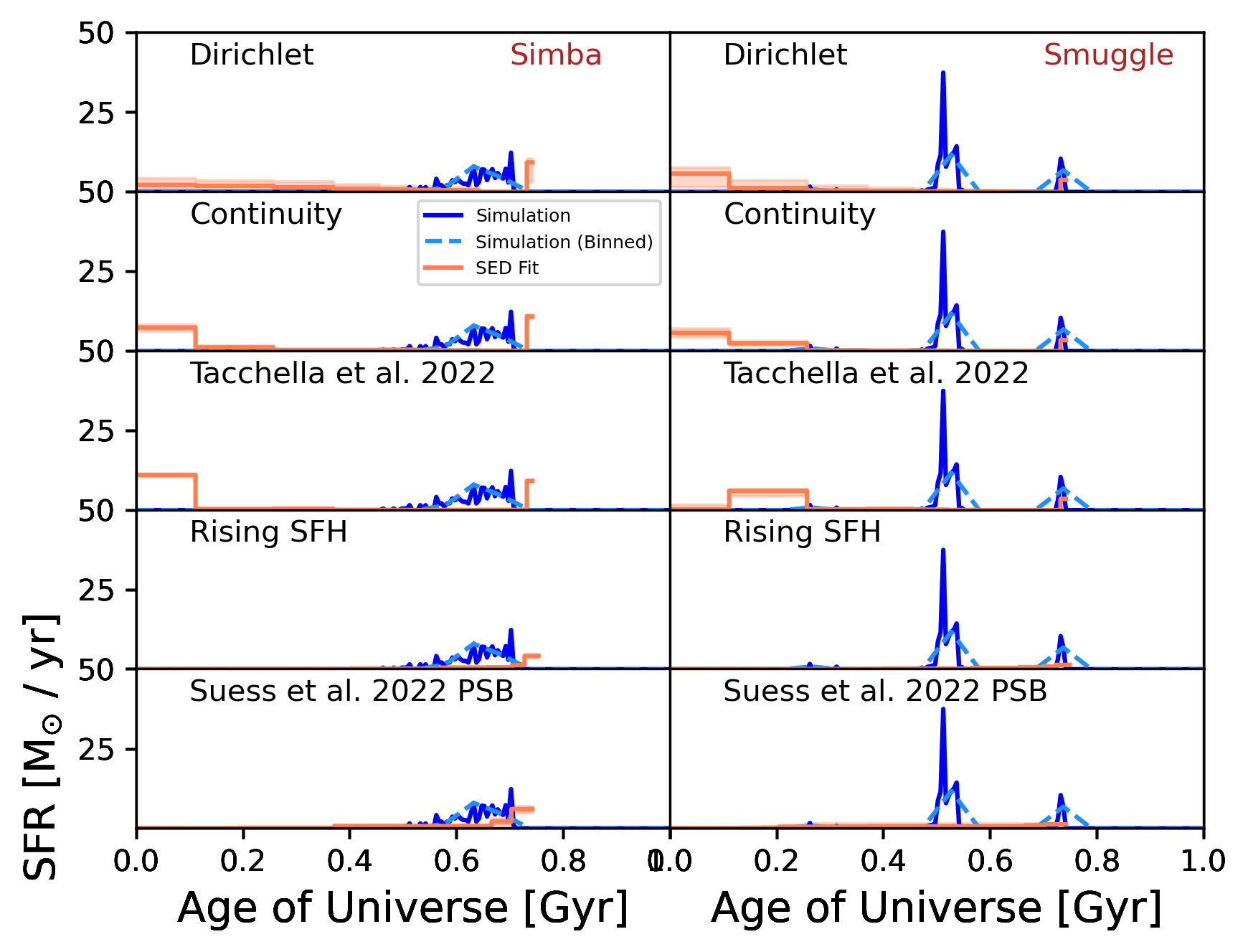}
    \caption{{\bf Star formation histories for the $5^{\rm th}$ most massive galaxy in the {\sc simba} (left) and {\sc smuggle} (right) simulation. }}
  \end{centering}
\end{figure*}

\begin{figure*}
  \begin{centering}
    \includegraphics{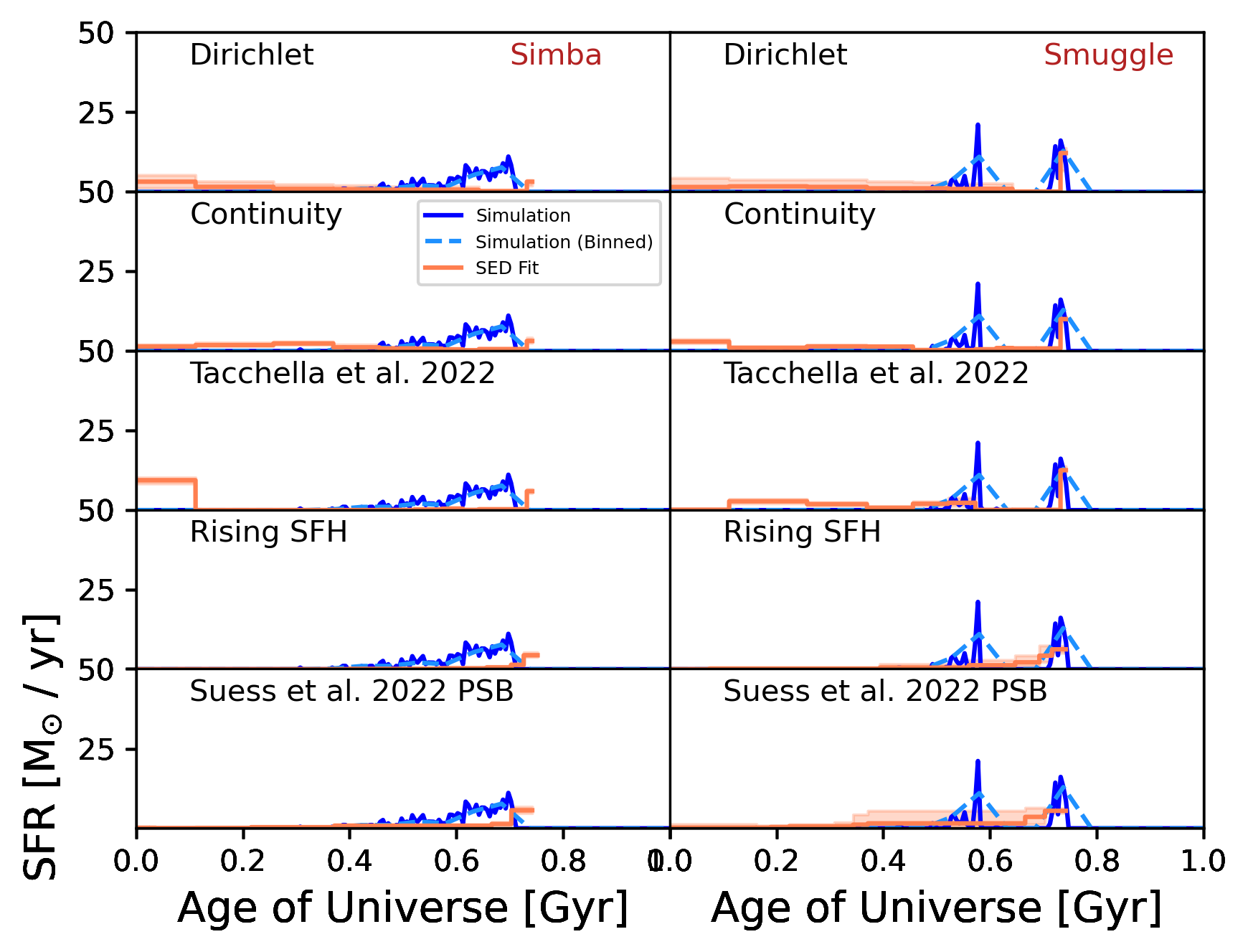}
    \caption{{\bf Star formation histories for the $6^{\rm th}$ most massive galaxy in the {\sc simba} (left) and {\sc smuggle} (right) simulation. }}
  \end{centering}
\end{figure*}

\begin{figure*}
  \begin{centering}
    \includegraphics{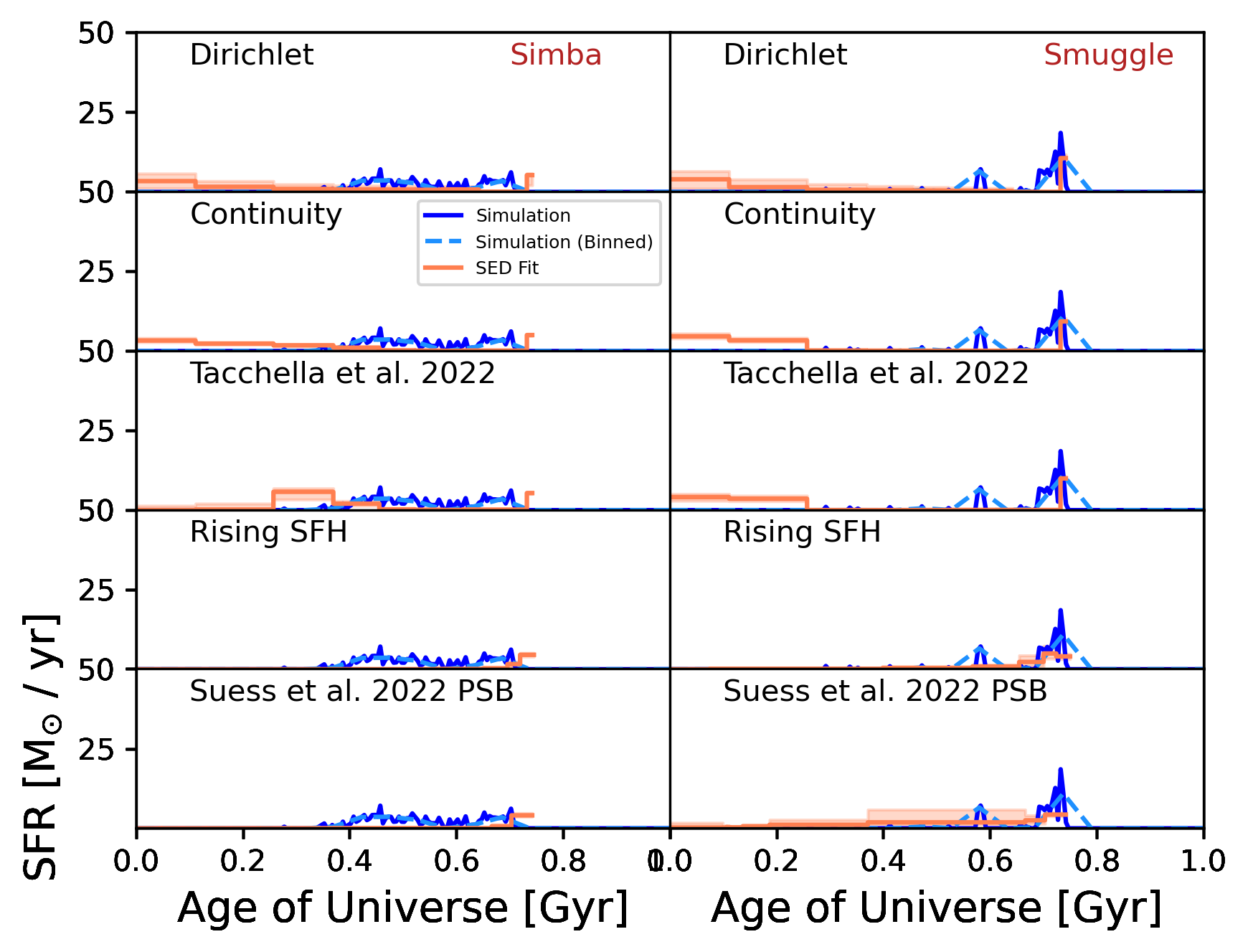}
    \caption{{\bf Star formation histories for the $7^{\rm th}$ most massive galaxy in the {\sc simba} (left) and {\sc smuggle} (right) simulation. }}
  \end{centering}
\end{figure*}

\begin{figure*}
  \begin{centering}
    \includegraphics{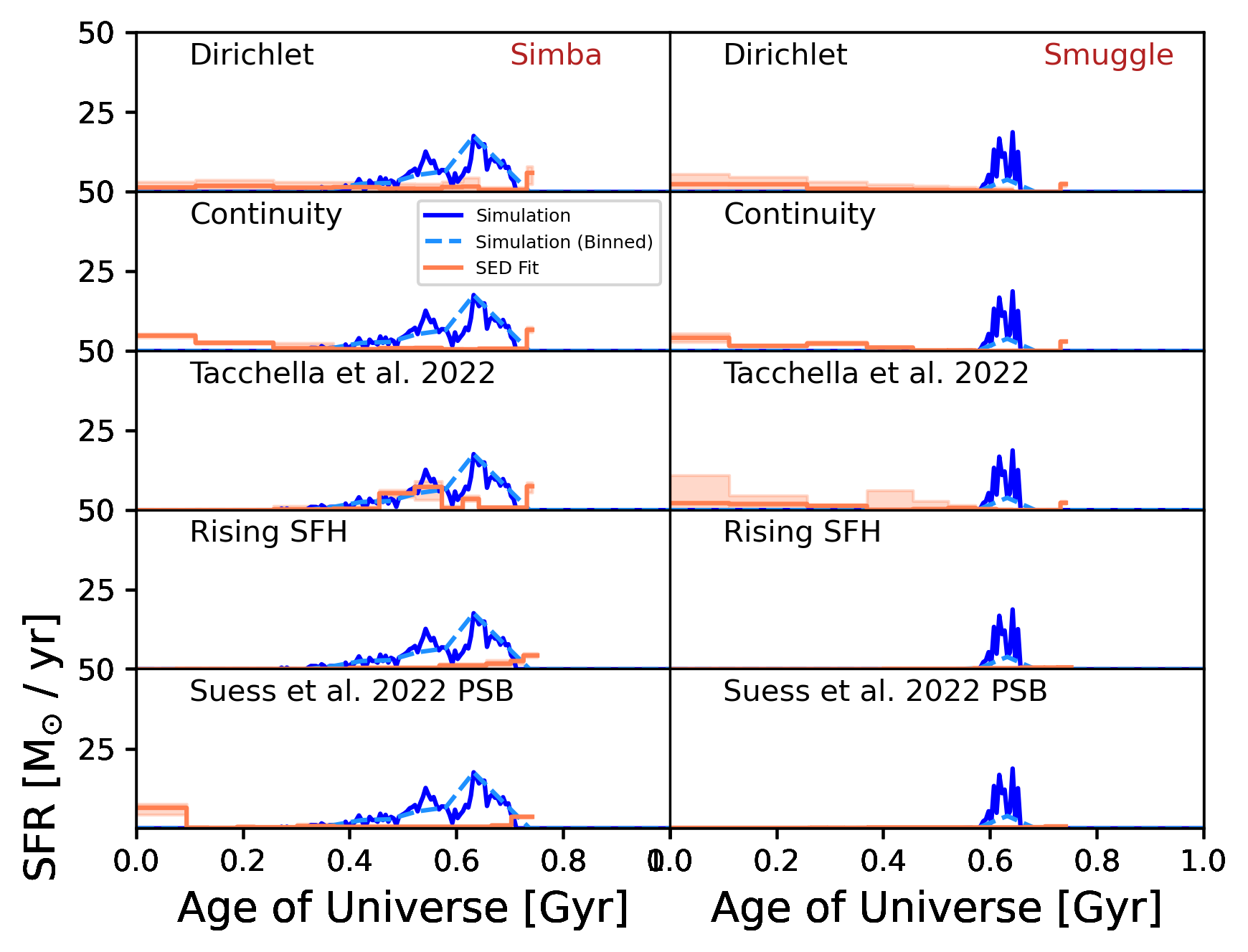}
    \caption{{\bf Star formation histories for the $8^{\rm th}$ most massive galaxy in the {\sc simba} (left) and {\sc smuggle} (right) simulation. }}
  \end{centering}
\end{figure*}

\begin{figure*}
  \begin{centering}
    \includegraphics{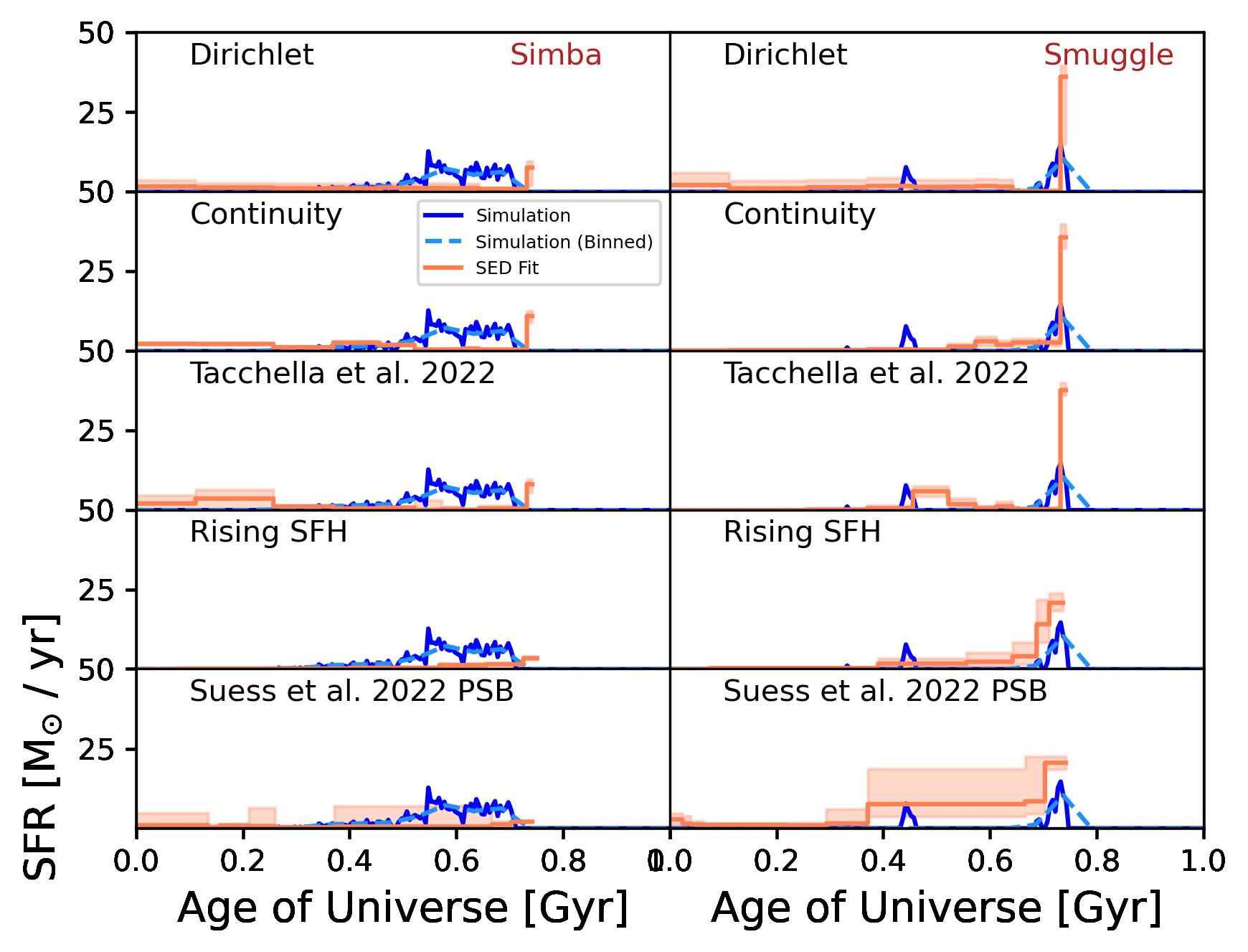}
    \caption{{\bf Star formation histories for the $9^{\rm th}$ most massive galaxy in the {\sc simba} (left) and {\sc smuggle} (right) simulation. }}
  \end{centering}
\end{figure*}

\begin{figure*}
  \begin{centering}
    \includegraphics{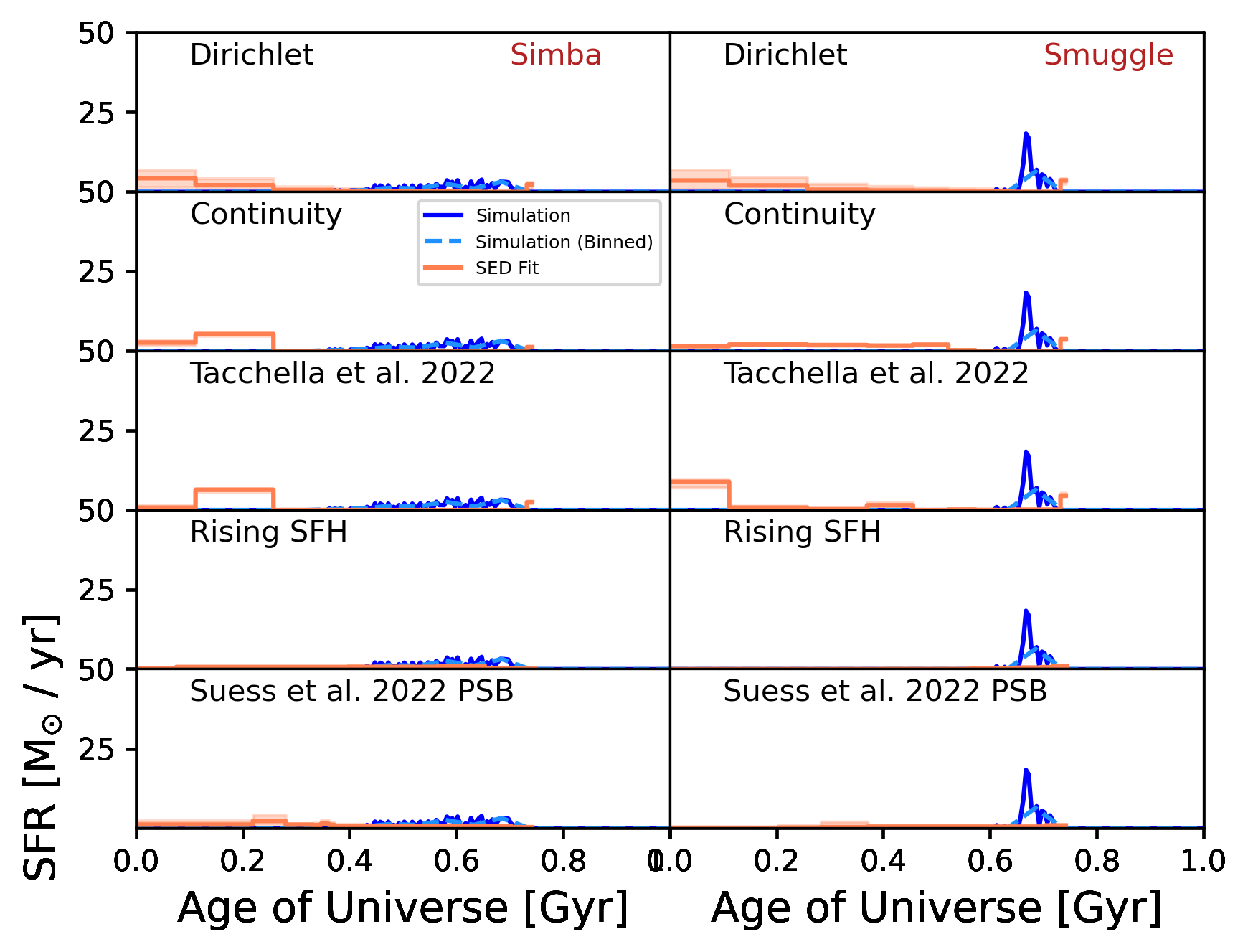}
    \caption{{\bf Star formation histories for the $10^{\rm th}$ most massive galaxy in the {\sc simba} (left) and {\sc smuggle} (right) simulation. }}
  \end{centering}
\end{figure*}

\begin{figure*}
  \begin{centering}
    \includegraphics{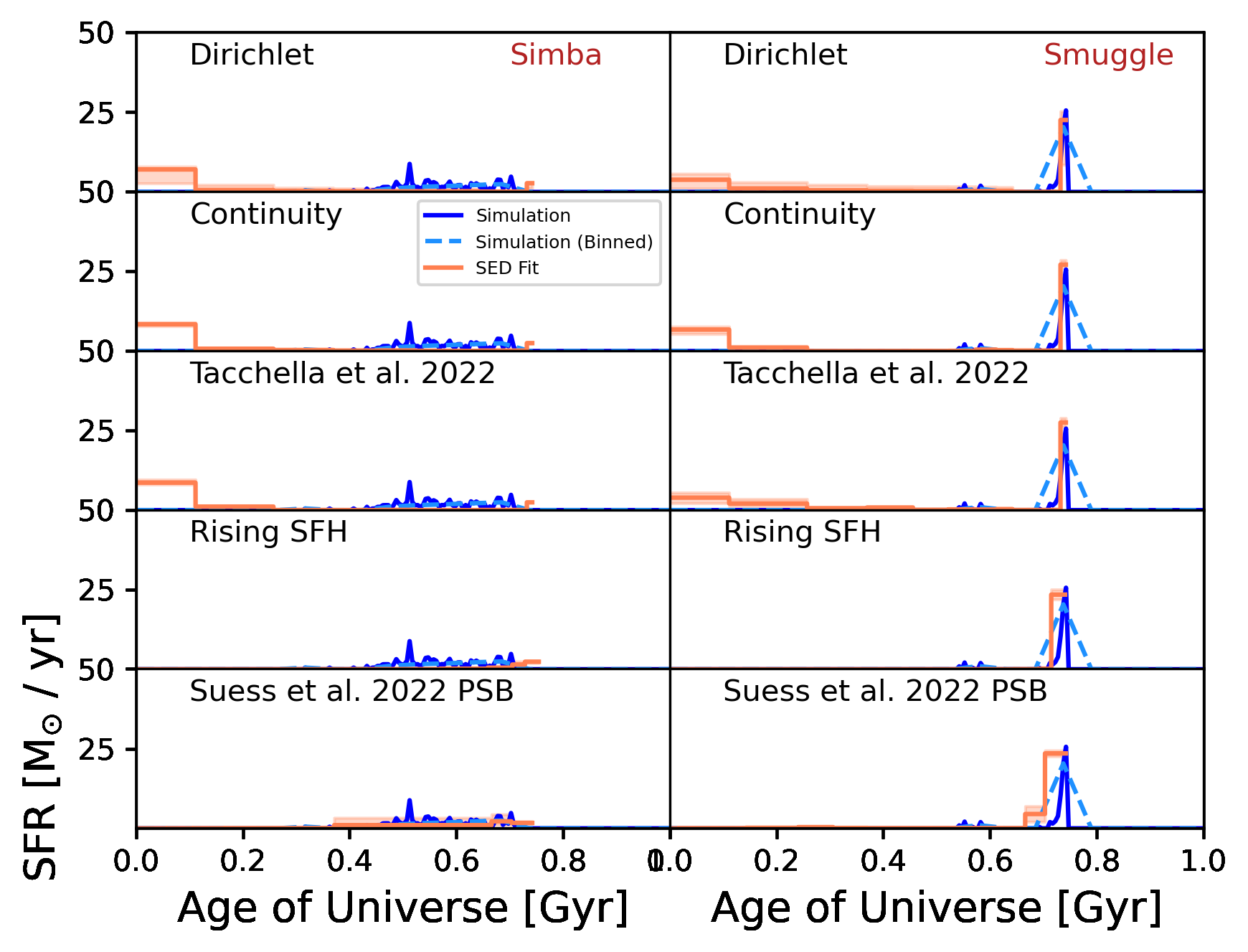}
    \caption{{\bf Star formation histories for the $11^{\rm th}$ most massive galaxy in the {\sc simba} (left) and {\sc smuggle} (right) simulation. }}
  \end{centering}
\end{figure*}

\begin{figure*}
  \begin{centering}
    \includegraphics{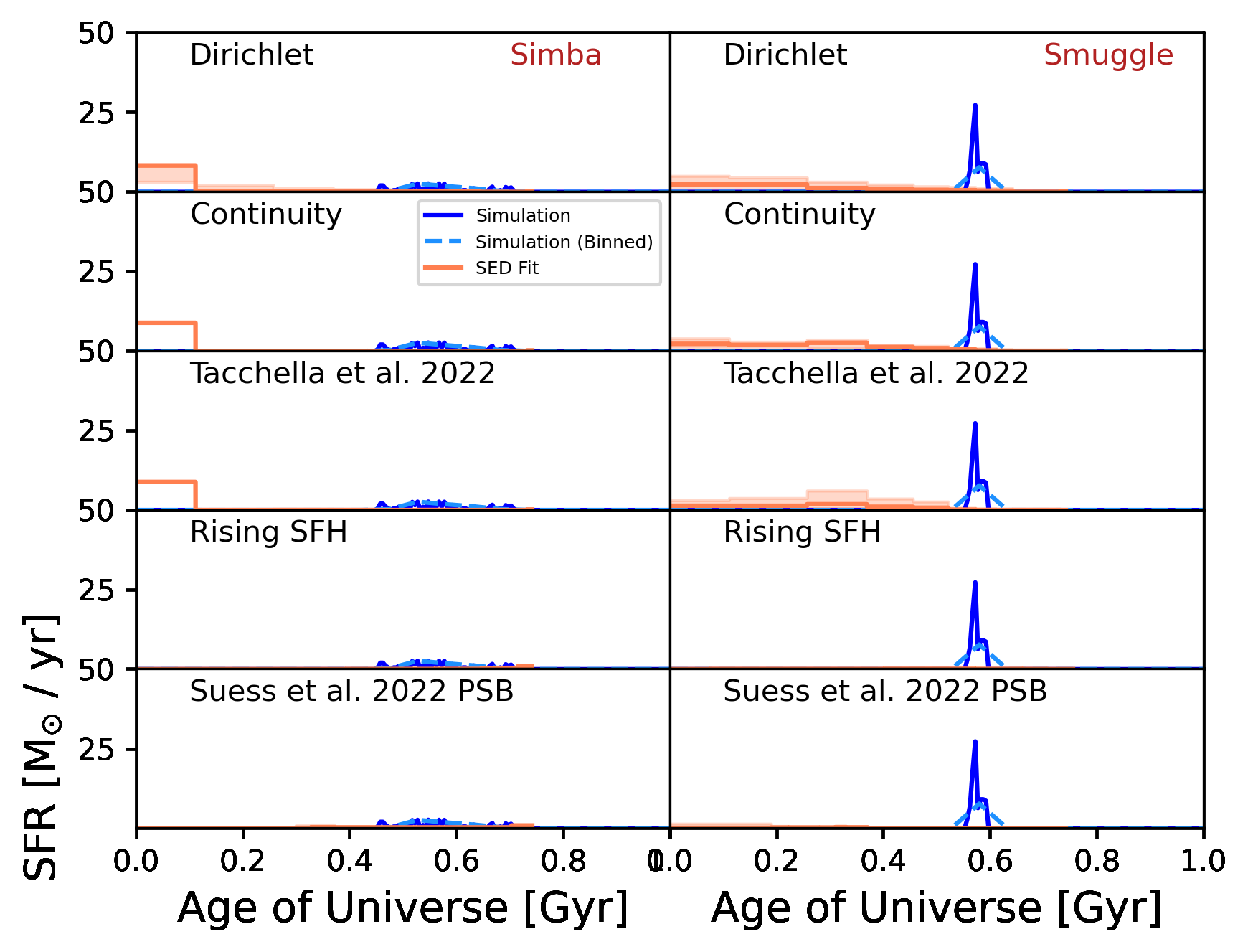}
    \caption{{\bf Star formation histories for the $12^{\rm th}$ most massive galaxy in the {\sc simba} (left) and {\sc smuggle} (right) simulation. }}
  \end{centering}
\end{figure*}

\begin{figure*}
  \begin{centering}
    \includegraphics{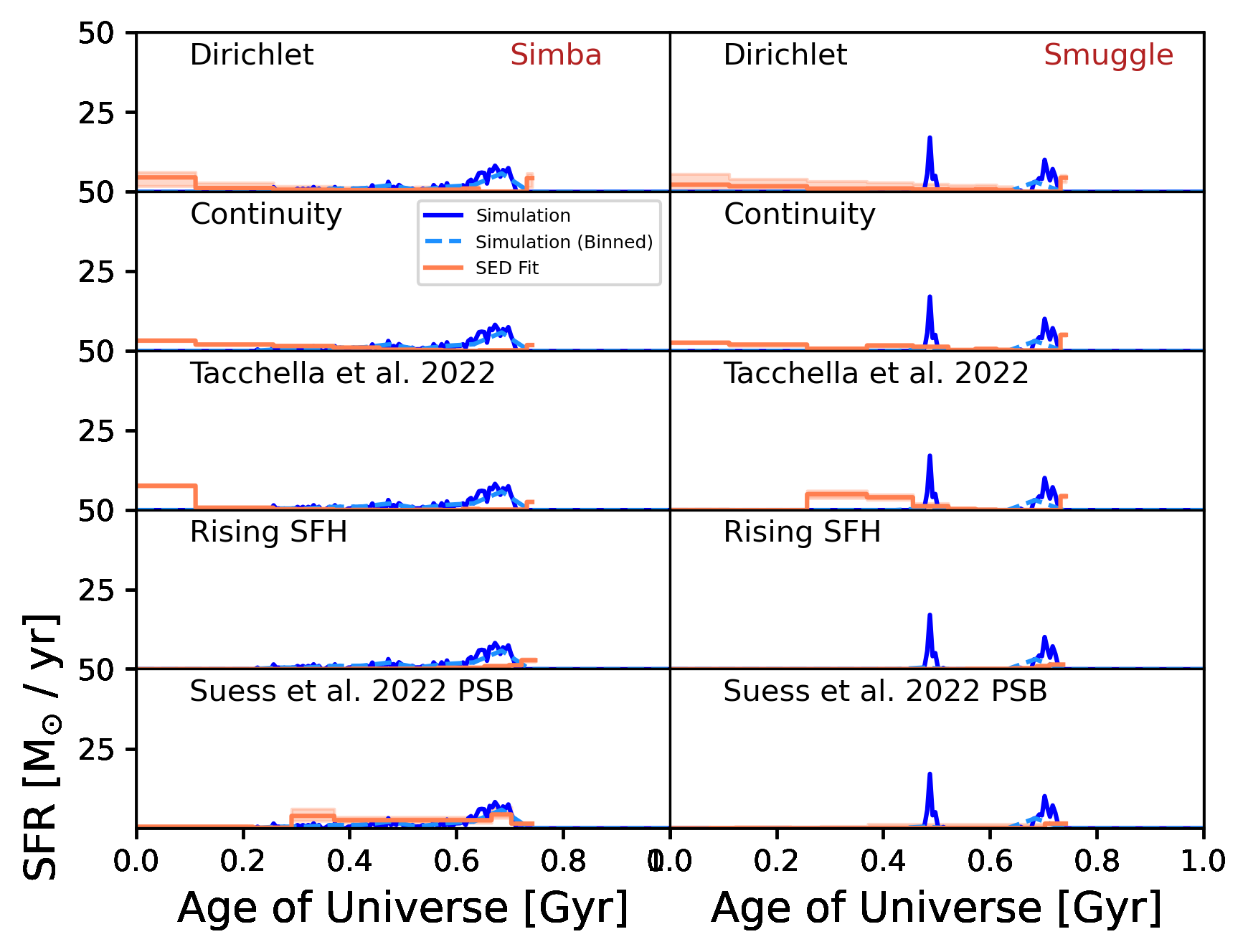}
    \caption{{\bf Star formation histories for the $13^{\rm th}$ most massive galaxy in the {\sc simba} (left) and {\sc smuggle} (right) simulation. }}
  \end{centering}
\end{figure*}

\begin{figure*}
  \begin{centering}
    \includegraphics{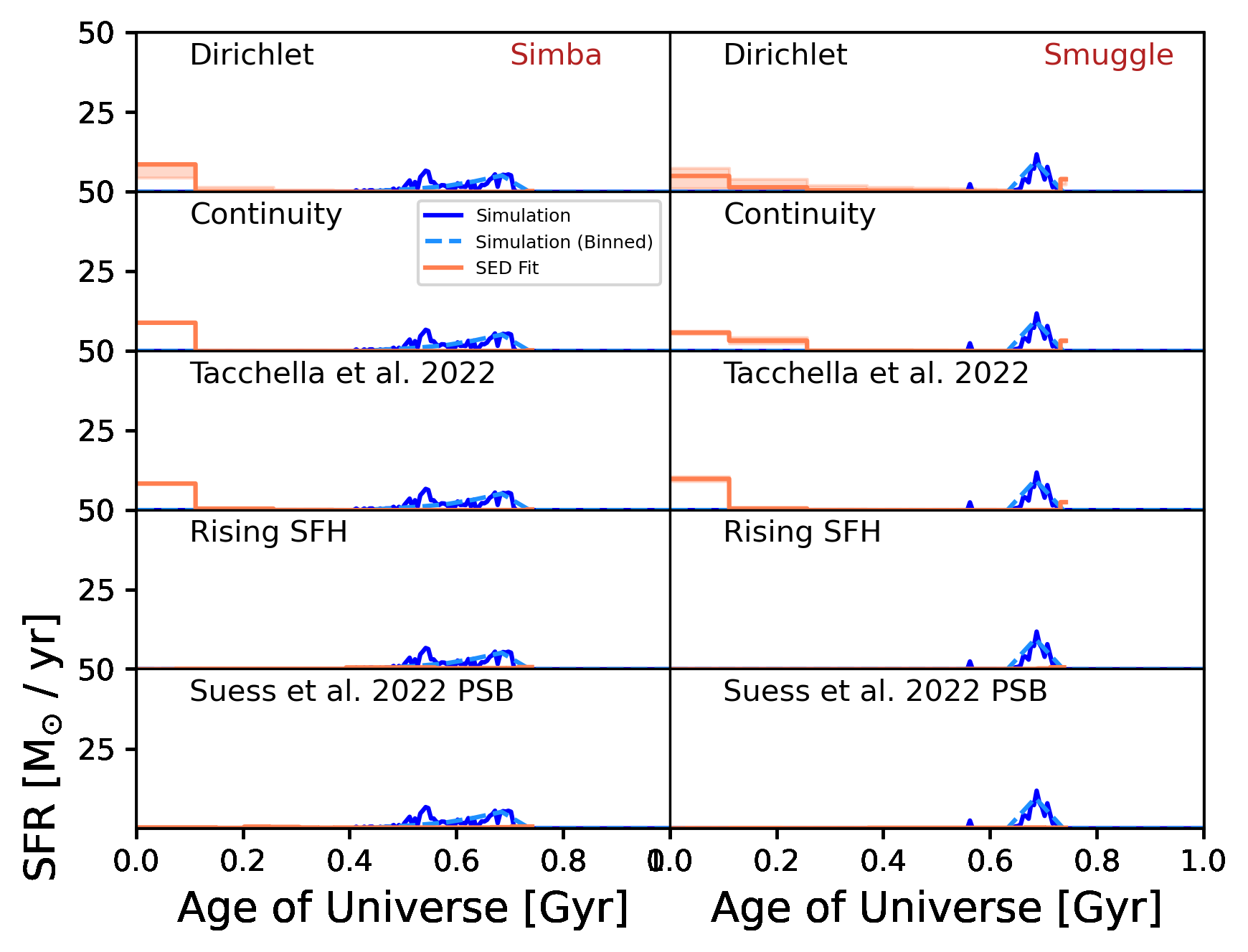}
    \caption{{\bf Star formation histories for the $14^{\rm th}$ most massive galaxy in the {\sc simba} (left) and {\sc smuggle} (right) simulation. }}
  \end{centering}
\end{figure*}

\begin{figure*}
  \begin{centering}
    \includegraphics{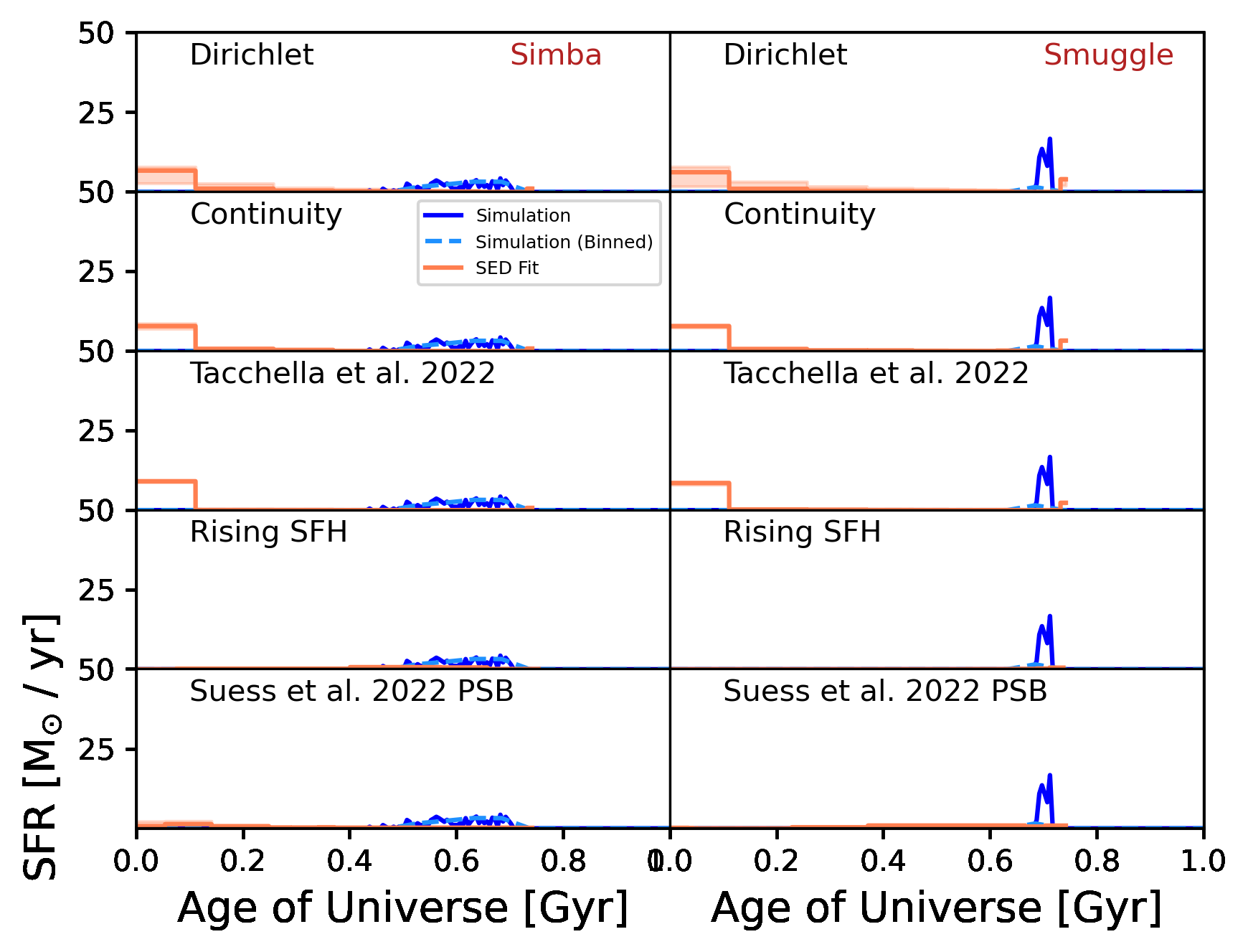}
    \caption{{\bf Star formation histories for the $15^{\rm th}$ most massive galaxy in the {\sc simba} (left) and {\sc smuggle} (right) simulation. }}
  \end{centering}
\end{figure*}

\begin{figure*}
  \begin{centering}
    \includegraphics{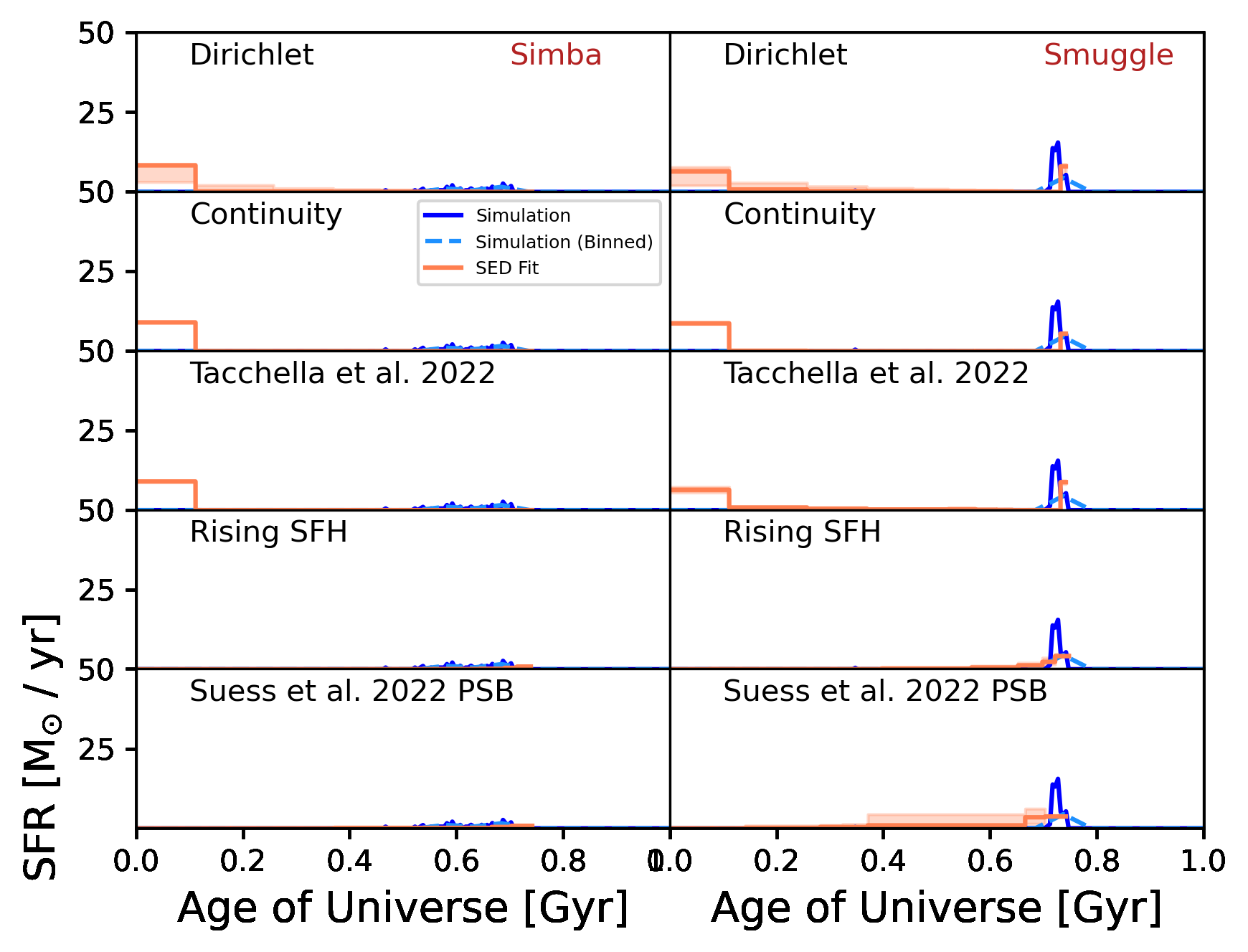}
    \caption{{\bf Star formation histories for the $16^{\rm th}$ most massive galaxy in the {\sc simba} (left) and {\sc smuggle} (right) simulation. }}
  \end{centering}
\end{figure*}

\begin{figure*}
  \begin{centering}
    \includegraphics{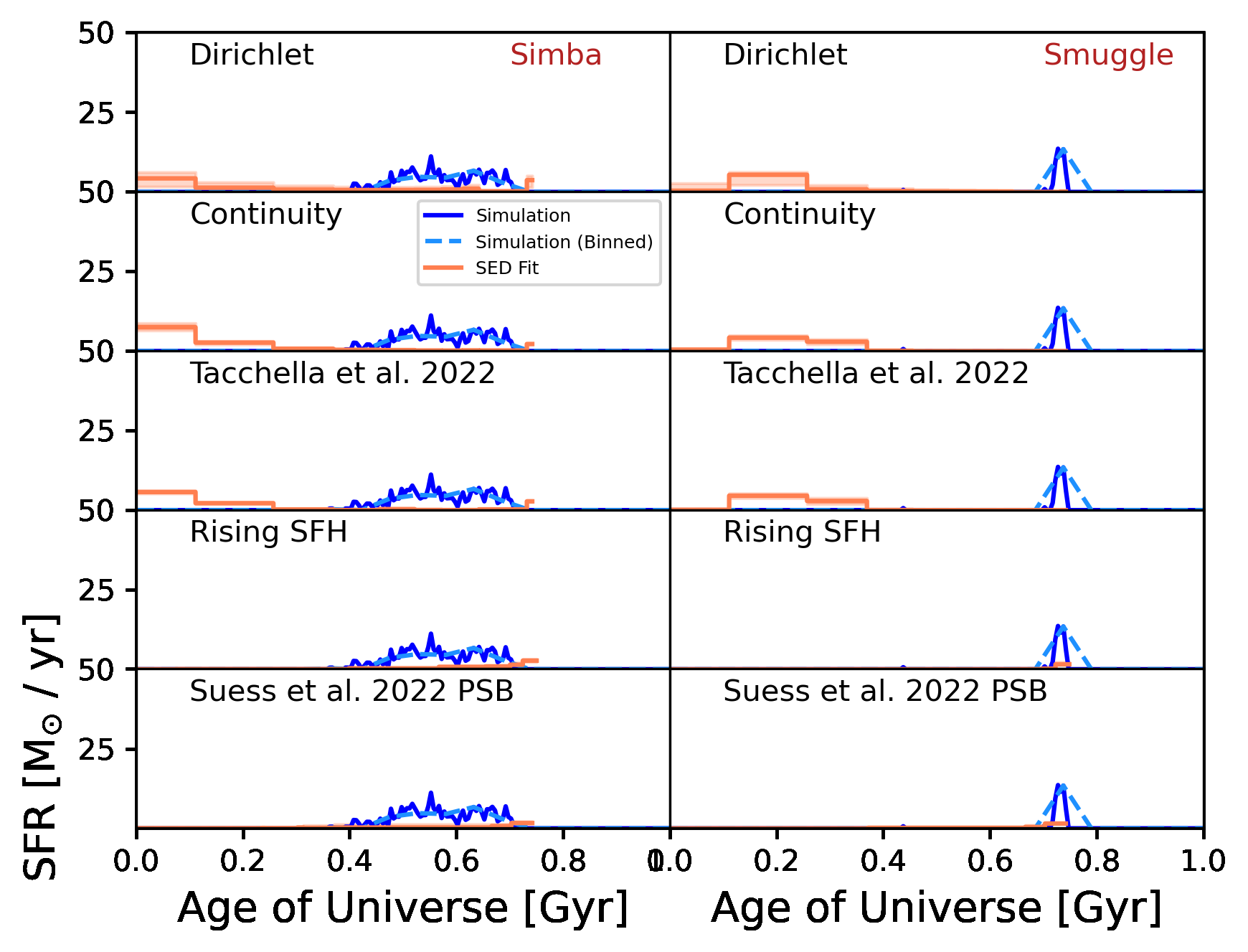}
    \caption{{\bf Star formation histories for the $17^{\rm th}$ most massive galaxy in the {\sc simba} (left) and {\sc smuggle} (right) simulation. }}
  \end{centering}
\end{figure*}

\begin{figure*}
  \begin{centering}
    \includegraphics{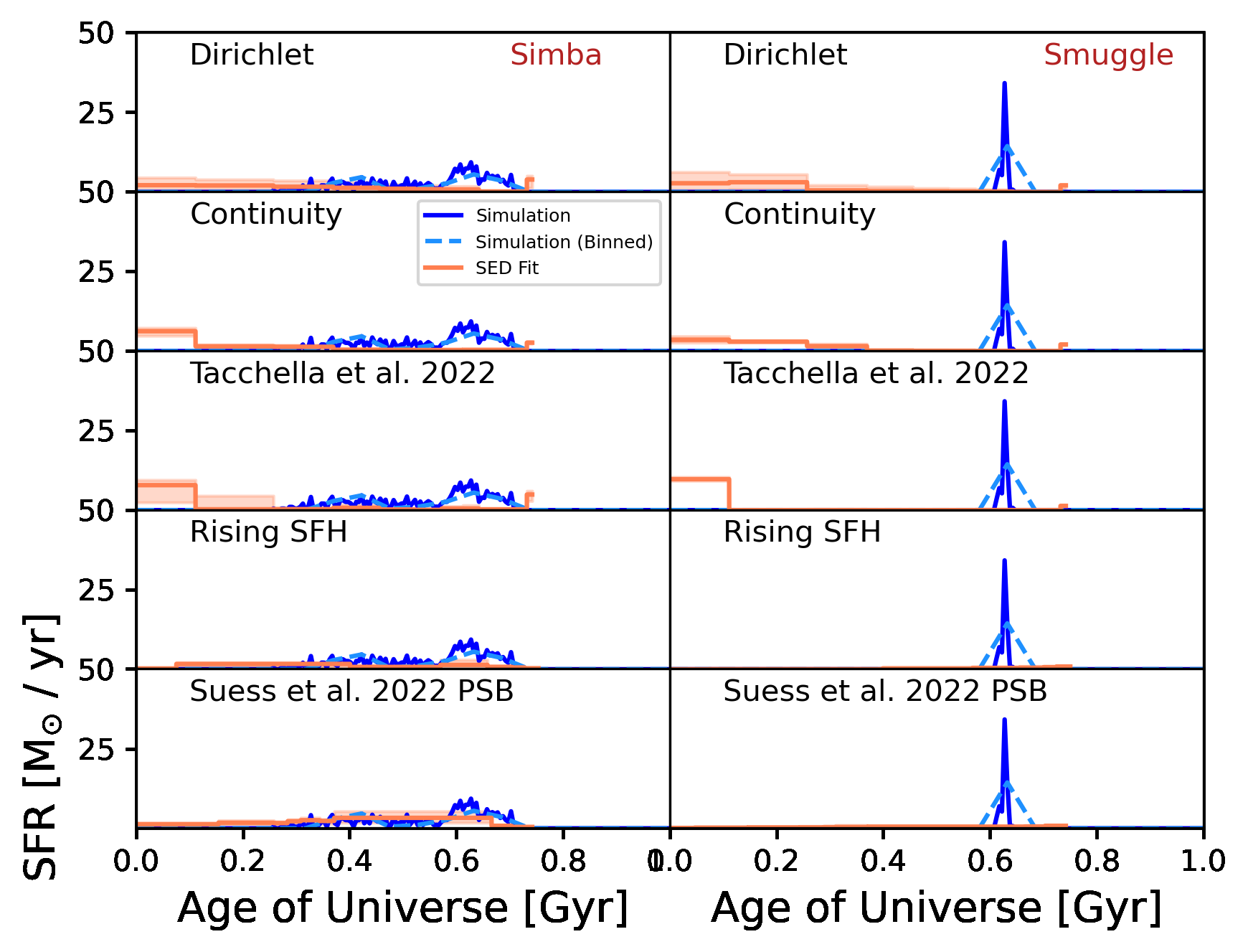}
    \caption{{\bf Star formation histories for the $18^{\rm th}$ most massive galaxy in the {\sc simba} (left) and {\sc smuggle} (right) simulation. }}
  \end{centering}
\end{figure*}

\begin{figure*}
  \begin{centering}
    \includegraphics{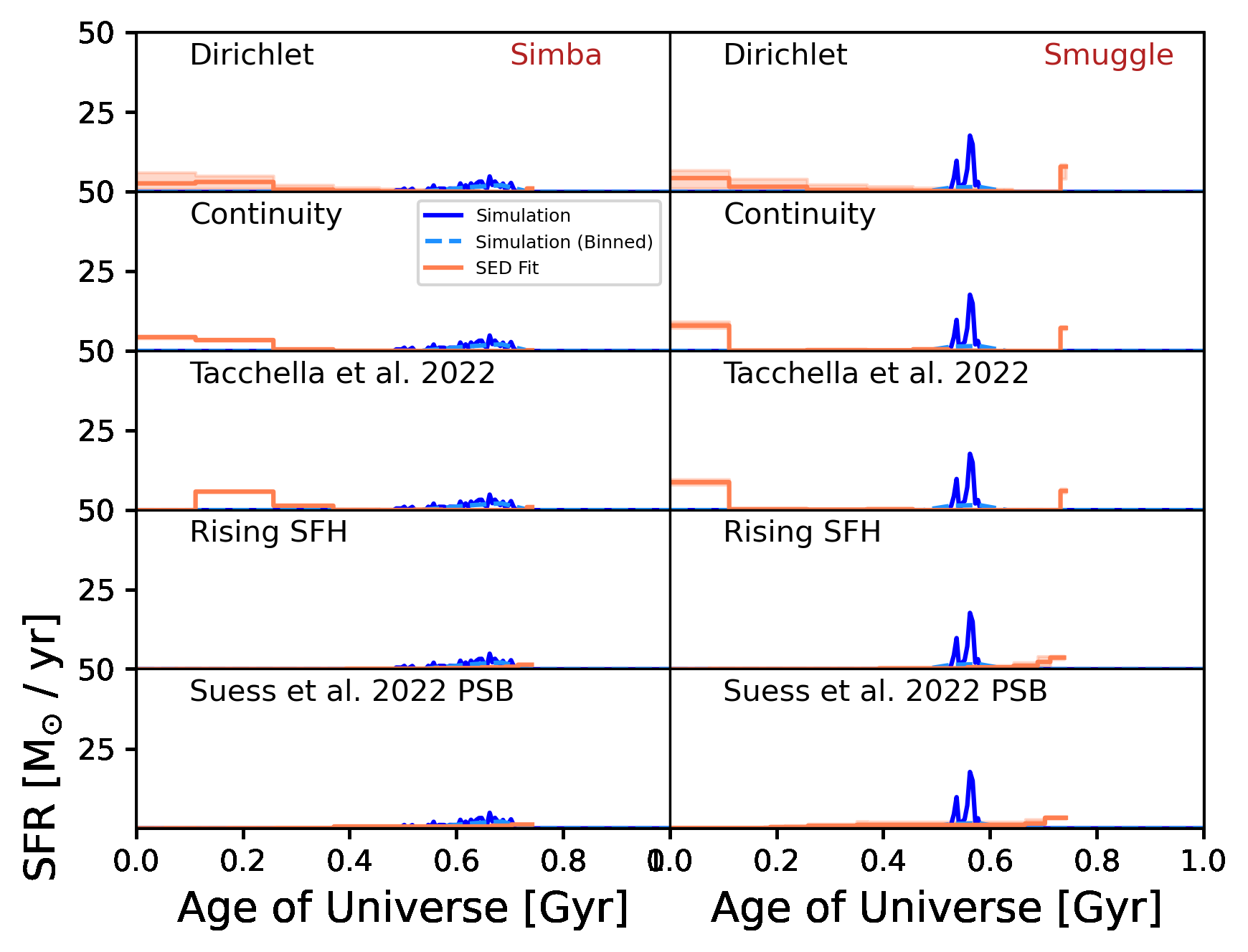}
    \caption{{\bf Star formation histories for the $19^{\rm th}$ most massive galaxy in the {\sc simba} (left) and {\sc smuggle} (right) simulation. }}
  \end{centering}
\end{figure*}

\begin{figure*}
  \begin{centering}
    \includegraphics{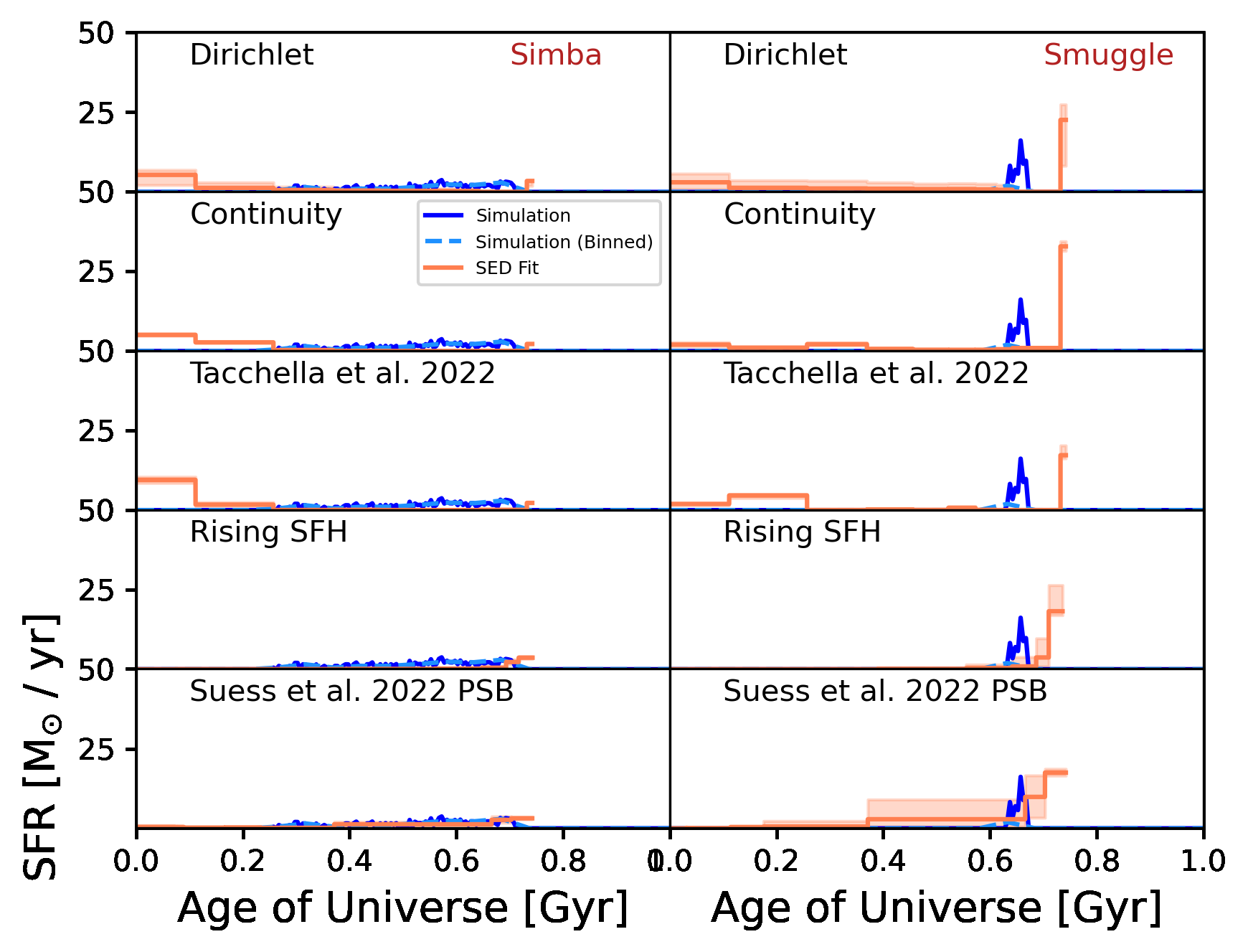}
    \caption{{\bf Star formation histories for the $20^{\rm th}$ most massive galaxy in the {\sc simba} (left) and {\sc smuggle} (right) simulation. }}
  \end{centering}
\end{figure*}

\end{document}